\documentclass[12pt,preprint]{aastex}

\shortauthors{Matthews et al.}
\shorttitle{HI Observations of X Her}

\begin{document}
\newcommand{\ang}{\rm \AA}
\newcommand{\msun}{M$_\odot$}
\newcommand{\lsun}{L$_\odot$}
\newcommand{\days}{$d$}
\newcommand{\degree}{$^\circ$}
\newcommand{\ud}{{\rm d}}
\newcommand{\as}[2]{$#1''\,\hspace{-1.7mm}.\hspace{.0mm}#2$}
\newcommand{\am}[2]{$#1'\,\hspace{-1.7mm}.\hspace{.0mm}#2$}
\newcommand{\ad}[2]{$#1^{\circ}\,\hspace{-1.7mm}.\hspace{.0mm}#2$}
\newcommand{\lsim}{~\rlap{$<$}{\lower 1.0ex\hbox{$\sim$}}}
\newcommand{\gsim}{~\rlap{$>$}{\lower 1.0ex\hbox{$\sim$}}}
\newcommand{\HA}{H$\alpha$}
\newcommand{\HII}{\mbox{H\,{\sc ii}}}
\newcommand{\kms}{\mbox{km s$^{-1}$}}
\newcommand{\HI}{\mbox{H\,{\sc i}}}
\newcommand{\KI}{\mbox{K\,{\sc i}}}
\newcommand{\nan}{Nan\c{c}ay}
\newcommand{\galex}{{\it GALEX}}
\newcommand{\jks}{Jy~km~s$^{-1}$}

\title{\HI\ Observations of the Asymptotic Giant Branch Star 
X Herculis: Discovery of an Extended Circumstellar Wake Superposed on 
a Compact High-Velocity Cloud}

\author{L. D. Matthews\altaffilmark{1},  
Y. Libert\altaffilmark{2},
E. G\'erard\altaffilmark{3}, T. Le~Bertre\altaffilmark{4}, 
M. C. Johnson\altaffilmark{5},
T. M. Dame\altaffilmark{6}}

\altaffiltext{1}{MIT Haystack Observatory, Off Route 40, Westford, MA
  01886}
\altaffiltext{2}{IRAM, 300 Rue de la Piscine, Domaine Universitaire,
  38406 Saint Martin d'H\`eres, France}
\altaffiltext{3}{GEPI, UMR 8111, Observatoire de Paris, 5 Place J
Janssen, F-92195 Meudon Cedex, France}
\altaffiltext{4}{LERMA, UMR 8112, Observatoire de Paris, 61 av.
de l'Observatoire, F-75014 Paris, France}
\altaffiltext{5}{Astronomy Department, Wesleyan University,
  Middletown, CT 06459}
\altaffiltext{6}{Harvard-Smithsonian Center for Astrophysics, 60
  Garden Street, MS-42, Cambridge, MA 02138}

\begin{abstract}
We report \HI\ 21-cm line observations of the asymptotic giant branch
(AGB) star X~Her obtained with the Robert C. Byrd Green Bank Telescope (GBT)
and the Very Large Array (VLA). We have unambiguously detected
\HI\ emission associated with the circumstellar envelope of the 
star, with a mass totaling
$M_{\rm HI}\approx 2.1\times10^{-3}~M_{\odot}$. The \HI\ distribution 
exhibits a  
head-tail morphology, similar to those previously observed around the
AGB stars
Mira and RS~Cnc. The tail
is elongated along the direction of the star's space motion, with a
total extent of $\gsim$\am{6}{0} (0.24~pc) in the plane of the sky. 
We also detect a systematic radial velocity gradient 
of $\sim$6.5~\kms\
across the \HI\ envelope. These results are
consistent with the \HI\ emission tracing a turbulent wake that arises
from the motion of a mass-losing star through the
interstellar medium (ISM). 
GBT mapping
of a $2^{\circ}\times2^{\circ}$ region around X~Her
reveals that the star lies (in projection) near
the periphery of a much larger \HI\ cloud that also exhibits signatures of
interaction with the ISM. The
properties of the cloud are consistent with those of compact 
high-velocity clouds. Using $^{12}$CO $J$=1-0 
observations, we have placed an upper limit on its molecular gas
content of $N_{H2}<1.3\times10^{20}$~cm$^{-2}$. Although the distance to the
cloud is poorly constrained, the probability of a chance 
coincidence in position, velocity, and apparent 
position angle of space motion between X~Her and the
cloud is extremely small, suggesting a possible physical
association. However, the  large \HI\ mass of the cloud
($\gsim2.4~M_{\odot}$) and the blueshift of 
its mean velocity relative to
X~Her are inconsistent with an origin tied directly to ejection from the star. 

\end{abstract}

\keywords{stars: AGB and
post-AGB -- stars: Individual (X Her) --- stars: winds, outflows -- 
radio lines: stars}  

\section{Introduction}
Stars on the asymptotic giant branch (AGB) undergo copious mass-loss
through cool, low-velocity winds
($V_{\rm outflow}\sim$10~\kms). 
The material dispersed by these winds supplies a
primary source of chemical enrichment to the interstellar medium
(ISM; e.g., Cristallo et al. 2009 and references therein).
In addition, the outflows and space motions of these mass-losing stars
play a role in shaping the structure of the ISM
on parsec and sub-parsec scales (Villaver et al. 2003; 
Wareing et al. 2007b). 
Furthermore, the distribution of circumstellar debris on the
largest scales may be important for the evolution of planetary nebulae
as well as some Type~Ia supernovae (e.g., Wang et al. 2004; Deng et
al. 2004).

It has long been recognized that the interstellar
environment of evolved stars can have a profound impact on the ultimate
mass, 
size, shape, and chemical composition of their circumstellar envelopes
(hereafter CSEs; 
e.g., Smith 1976; Isaacman 1979; Serabyn et al. 1991; 
Young et al. 1993; Zijlstra \& Weinberger 2002; Villaver
et al. 2002, 2003; Wareing et
al. 2006b; 2007a, b, c; see also the review by Stencel 2009). 
Nonetheless, this 
has been largely ignored in many theoretical investigations of AGB star
evolution, which still treat such stars as evolving in isolation.
Moreover,
the interfaces through which AGB stars interact with their environments have
remained poorly studied observationally. 
One problem is that CSEs
can be enormously extended ($\gsim$1~pc), and
their chemical composition changes as a
function of distance from the star, as densities drop and 
molecules become dissociated by
the interstellar radiation field or other radiation sources. 
The result is that the some of the most frequently used observational 
tracers of CSEs
(e.g., CO; SiO, H$_{2}$O, and OH masers) 
do not probe the outermost CSE or its interaction zone
with the ISM. In some cases, more extended material can be traced via
imaging observations in the infrared (Zijlstra \& Weinberger
2002; Ueta et
al. 2006, 2010a, b; Ladjal et al. 2010) or in the
far-ultraviolet (Martin et al. 2007; Sahai \& Chronopoulos 2010),
but such data do not supply any direct kinematic information.

Our recent work has demonstrated that the \HI\ 21-cm line of neutral
hydrogen is a powerful tool for the study
of extended CSEs
and their interface with their larger-scale environments 
(e.g., G\'erard \& Le~Bertre 2006; 
Matthews \& Reid 2007; Libert
et al. 2007, 2008, 2010a,b; Matthews et
al. 2008).  
Not only is hydrogen the most plentiful constituent of CSEs, 
but \HI\ line measurements provide kinematic information
and can be directly translated into \HI\ mass
measurements. Furthermore,  because \HI\ is not destroyed by the interstellar
radiation field, \HI\ measurements are particularly well-suited to probing the
outermost reaches of CSEs.

In this paper, we continue our investigation of the \HI\ properties of
evolved stars with a study of the nearby, semi-regular
variable star X~Herculis (X~Her). For the first time, we combine 
Very Large Array (VLA) \HI\ imaging
observations with \HI\ mapping 
obtained with the Robert C. Byrd Green Bank Telescope (GBT). 
In concert, these observations allow us to
probe the CSE and its environs on spatial
scales ranging from $\sim1'$ to $\sim1^{\circ}$. At the smaller scales,
our observations
complement previous studies of the inner CSE from CO and SiO
line observations.  On large scales, our observations show
that X~Her overlaps in velocity and projected position with a
compact high velocity cloud. We investigate the properties of this
cloud in detail 
and evaluate the possibility that it could be associated with X~Her.

\section{An Overview of X Her}
\subsection{Stellar Properties\protect\label{properties}}
X~Her is an oxygen-rich, semi-regular variable star of type SRb and 
a spectral classification that ranges in the literature from M6 to
M8. Its rather large velocity relative to the local standard of rest 
($V_{\rm LSR}\approx -73$~\kms; Knapp et
al. 1998) makes it well-suited for \HI\ observations, since it is
well-separated from the bulk of the Galactic emission along the line-of-sight.
The star has been found to have both short and
long pulsation cycles; periods of 102 and 178~days have been reported by 
Kiss et al. (1999) based on photometric measurements, 
while Hinkle et al. (2002) found a longer period of
658~days based on a spectroscopic analysis. The 
mass and luminosity of X~Her are estimated to be $1.9~M_{\odot}$ and
3570~$L_{\odot}$, respectively (Dyck et al. 1998; Dumm \& Schild 1998). 
Some additional stellar properties
of X~Her are summarized in Tables~1 \& 2. Throughout this work we
adopt a distance of $d_{*}\approx$140~pc based on the {\it Hipparcos}
parallax measurement of 7.30$\pm$0.40~mas (van Leeuwen 2007). 

\begin{deluxetable}{lcc}
\tabletypesize{\scriptsize}
\tablewidth{0pc}
\tablenum{1}
\tablecaption{Coordinates and Stellar Properties of X Herculis}
\tablehead{\colhead{Parameter} & \colhead{Value} & \colhead{Ref.}}
\startdata

$\alpha$ (J2000.0) & 16 02 39.2 & 1\\

$\delta$ (J2000.0) & +47 14 25.3  & 1\\

$l$ & \ad{74}{46} & 1 \\

$b$ & \ad{+47}{78} & 1\\

Distance &  140~pc & 2\\

Chemistry & Oxygen-rich & 3\\

Variable Class & SRb & 3 \\

Spectral Type & M6-M8 & 1,4\\

Periods & 102$\pm$5, 178$\pm$5~days, 658.3$\pm$17.0~days & 4,5  \\

$T_{\rm eff}$ & 3281$\pm$130~K & 6\\

Radius & 183$\pm$4 $R_{\odot}$ & 6\\

Luminosity & 3570 $L_{\odot}$ & 6\\

Mass & 1.9~$M_{\odot}$ & 7\\


${\dot M}$\tablenotemark{a} 
& (0.34,1.1)$\times10^{-7}~M_{\odot}$ yr$^{-1}$ & 8\\

$V_{\rm outflow}$ & 3.5 \& 9~\kms\ & 8\\

\enddata

\tablecomments{Units of right ascension are hours, minutes, and
seconds, and units of declination are degrees, arcminutes, and
arcseconds. All quantities have been scaled to the distance adopted in
this paper.}

\tablenotetext{a}{Mass-loss rate based on two-component
CO line profiles.}

\tablerefs{(1) SIMBAD database; (2) van Leeuwen 2007; (3) Loup et
  al. 1993; (4) Hinkle et al. 2002; (5) Kiss et
  al. 1999.
(6) Dyck et al. 1998; (7) Dumm \& Schild 1998; (8) Knapp et al. 1998.  }

\end{deluxetable}
\begin{deluxetable}{lllll}
\tabletypesize{\scriptsize}
\tablewidth{0pc}
\tablenum{2}
\tablecaption{Previous Radial Velocity 
Measurements and Mass-Loss Estimates for
X Her}
\tablehead{\colhead{Line} & \colhead{$V_{\rm LSR}$ (\kms)} &
    \colhead{$V_{\rm out}$ (\kms)}  & \colhead{${\dot M}$ ($M_{\odot}$
      yr$^{-1}$)} & \colhead{Ref.}}
\startdata

\KI\ & $-80$ & ... & ... & 1\\
\\

Infrared\tablenotemark{a} & $-72.8$ & ... & ... & 2 \\
\\

SiO(2-1)\tablenotemark{b} & $-72.0$ & 2.5 
&$4.0\times10^{-8}$ &  3 \\
SiO(2-1)\tablenotemark{b} & $-73.0$ & 6.5
&$1.5\times10^{-7}$ &  3\\
\\

SiO(5-4)\tablenotemark{c} & ... & 4.9 
& ... &  4 \\
\\

CO(1-0) & $-71.0$ & 6.5 & $1.5\times10^{-7}$ &  5\\
\\

CO(2-1) &  $-73.2\pm0.4$ & 3.2$\pm0.5$ & $3.4\times10^{-8}$ & 6\\
CO(2-1) & $-72.8\pm0.8$ &8.5$\pm$1.0 & $1.1\times10^{-7}$ & 6\\ 
\\

CO(3-2) & $-73.1\pm0.3$ & 3.5$\pm1.4$ & $3.4\times10^{-8}$ & 6\\
CO(3-2) & $-73.2\pm0.5$ & 9.0$\pm1.0$ & $1.1\times10^{-7}$  & 6 \\ 
\\

\HI\ & $-70.6$ & 2\tablenotemark{d} & $\sim10^{-7}$ & 7 \\
\HI\ &$-72.2$ & 6.5\tablenotemark{d} & $\sim10^{-7}$ & 7\\ 

\enddata

\tablecomments{ 
Explanation of columns: (1) spectral line used for the
  measurement; (2) radial velocity relative to the local standard of
  rest (LSR); (3) outflow velocity; (4) derived mass-loss rate in
  solar masses per year; (5)
  reference. Two sets of parameters are quoted for a given spectral
  line in cases where the
  line profile was found to be complex. }

\tablenotetext{a}{A cross-correlation of multiple infrared
  lines was used; velocity was translated from
  the heliocentric to the LSR reference frame using $V_{\rm hel} -
  V_{\rm LSR}=-17.5$~\kms.}

\tablenotetext{b}{Based on the thermal SiO $v$=0, $J$=2-1 line.}

\tablenotetext{c}{Based on the SiO $v$=0, $J$=5-4 line.}

\tablenotetext{d}{Expansion velocities are estimated based on
  the half-width at half maximum of the line profile.}

\tablerefs{(1) Wallerstein \& Dominy 1988; 
(2) Hinkle et al. 2002; (3) Gonz\'alez Delgado et
  al. 2003; (4) Sahai \& Wannier 1992; 
(5) Kerschbaum et al. 1996; (6) Knapp et al. 1998; 
  (7) Gardan et al. 2006. }

\end{deluxetable}

\subsection{Previous Studies of the X~Her CSE\protect\label{oldstuff}}
One of the first detailed studies of the CSE of X~Her was by Kahane \& Jura
(1996), who used the IRAM 30-m telescope to map the circumstellar CO(1-0) and
CO(2-1) emission. The authors found the CO line profile shapes to be
more complex than those seen in many other AGB
stars, comprising both broad and narrow components whose spectral widths
correspond to outflow velocities of $\sim$10~\kms\ and $\sim$3~\kms,
respectively (see also Knapp et al. 1998). 
Kahane \& Jura were also able to  marginally resolve the CSE spatially
in their CO(2-1) observations (with a beam FWHM of $12''$);
based on these data, they argued for the existence of
a multi-component wind, including a weakly collimated
bipolar outflow. 

Subsequent higher resolution
CO(1-0) imaging observations by Nakashima (2005; 
FWHM$\approx3''$) confirmed the
presence of a bipolar outflow associated with the broader line
component of X~Her. This outflow extends to $\sim\pm10''$
($\pm$1400~AU) from the star. Nakashima suggested that the narrower CO
line component might
arise either from a second, smaller-scale bipolar outflow, or from a 
disk that is rotating or  expanding. More recent CO imaging
observations by Castro-Carrizo et al. (2010) with 1$''$-2$''$
resolution support the latter
interpretation---i.e., an expanding, equatorial disk that lies
perpendicular to the larger-scale bipolar flow. 

The presence of these 
pronounced deviations from spherical symmetry already during the AGB phase
suggests the possibility that X~Her may be a binary (Kahane \&
Jura 1996). However,
to date, both speckle interferometry (Lu et al. 1987) and 
radial velocity searches have failed to provide any
evidence for a companion. Radial
velocity variations are present, but appear to be due
to the star's
pulsation (Hinkle et al. 2002). 

In 2006, Gardan et al. published the first \HI\ 21-cm line detection of 
X~Her based on observations with the
Nan\c{c}ay Radio Telescope (NRT).  The \HI\ line profile obtained by
these authors
appeared to be composite, comprising both broad
(FWHM$\sim$13~\kms) and narrow (FWHM$\sim$4~\kms) components. 
However, the linewidths and velocity centroids of these 
components differed from the broad and narrow
line components seen in CO and SiO (Table~2), suggesting they were not
obviously related.
Based on
NRT mapping, Gardan et al. found evidence  that the broad \HI\
component arose from material extending over $\gsim10'$ and that was
asymmetrically distributed about the
star, with a concentration to the northeast. Meanwhile, the narrow
component appeared to be unresolved ($<4'$) and centered on the star.
The total mass of circumstellar 
hydrogen inferred from 
the NRT study was $M_{\rm HI}\approx6.5\times10^{-3}~M_{\odot}$.

Gardan et al. (2006) suggested that the broad \HI\ line component that
they observed
might be explained by asymmetric mass loss, with an outflow along a
preferred direction. The implications of this would be significant,
since it would indicate the existence of pronounced large-scale asymmetries in the
circumstellar ejecta during the AGB stage---in stark contrast to the
commonly held picture of spherical mass-loss (cf. 
Huggins et al. 2009). Further, this would
suggest that the circumstellar 
atomic hydrogen is largely decoupled from the axisymmetric morphology
of the CSE traced on
smaller scales via CO emission.

Unfortunately, the limited resolution of the
NRT beam [$4'$~(E-W)$\times 22'$~(N-S)] permits only a coarse
characterization of the distribution of \HI\ 
emission around X~Her and its relationship to structures in the CSE on
smaller scales.
Furthermore, the need to decompose the circumstellar signal from
contaminating interstellar emission  
introduces uncertainties in the line decomposition and 
derived \HI\ fluxes. 
For these reasons, we have
obtained new, higher spatial 
resolution \HI\ observations of X~Her  and its environs 
using a combination of the VLA
and the GBT. In this paper, we describe our new and improved characterization
of the large-scale properties of the X~Her's CSE  that have
resulted from these observations. In addition, we investigate a new
puzzle uncovered by our observations---namely the relationship
between the CSE of X~Her and an adjacent ``high-velocity'' 
\HI\ cloud.

\section{VLA Observations\protect\label{VLAobs}}
X~Her was observed in the \HI\ 21-cm line with the VLA of
the National Radio Astronomy Observatory (NRAO)\footnote{The National
Radio Astronomy Observatory is operated by Associated Universities,
Inc., under cooperative agreement with the National Science
Foundation.} on 2007 May~5 
and 2007 May~10 using the most compact (D)
configuration (0.035-1.0~km baselines). This provided
sensitivity to emission on scales of up to $\sim$15$'$. The
primary beam of the VLA at our observing
frequency of 1420.3~MHz is $\sim 31'$. 

The VLA correlator was used in dual polarization (2AC) 
mode with a 0.78~MHz bandpass, yielding 256 spectral
channels with 3.05~kHz ($\sim$0.64~\kms) spacing. The band was centered
at a velocity of $-100$~\kms\ relative to the local standard of rest
(LSR); the band center was offset slightly from
the systemic velocity of the star ($V_{\rm sys,LSR}\approx-73$~\kms;
see Table~2)
to shift strong Galactic emission away from the band edge.

Observations of X~Her were interspersed with observations of the phase
calibrator, 1625+415, approximately every 20
minutes. 3C286 (1331+305) was used as a flux calibrator, and an
additional strong point source (1411+522) was observed as a bandpass 
calibrator. To 
insure that the absolute flux scale and bandpass calibration were not
corrupted by Galactic emission in the band, the
flux and bandpass calibrators were each observed twice,
first with
the band shifted by $+$1~MHz and then by $-1$~MHz,
relative to the band center used
for the observations of X~Her and 1625+415. 1625+415 was
also observed once at each of these offset frequencies to
allow more accurate bootstrapping of the absolute 
flux scale  to the X~Her data. The resulting flux scale has an
uncertainty of
$\sim$5\% (see below).
In total, $\sim$6.8 hours of integration were obtained on
X~Her. Approximately 6\% of the observed visibilities were
flagged because of radio frequency interference (RFI) or hardware problems. 

The VLA data were calibrated and reduced
using the Astronomical Image Processing System
(AIPS). At the time of our
observations, the VLA contained 26 operating antennas, 9 of which had
been retrofitted as part of the Expanded Very Large Array (EVLA)
upgrade. The use of this ``hybrid'' array necessitated special care in
the data reduction process. 

Traditionally, the calibration of VLA data sets make use of a
``channel 0'' data set formed by taking a vector average of the inner
75\% of the observing band. However, the
mismatch in bandpass response between the VLA and EVLA
antennas  
causes closure errors in the default channel~0 computed this
way.\footnote{http://www.vla.nrao.edu/astro/guides/evlareturn/postproc/}
Furthermore, the hardware used to convert the digital signals from the
EVLA antennas into analog signals for the VLA correlator causes
aliased power in the bottom 0.5~MHz of the baseband (i.e., at low-numbered
channels).\footnote{http://www.vla.nrao.edu/astro/guides/evlareturn/aliasing/}
This aliasing is of consequence because of the narrow bandwidth used for
our observations. However, it affects EVLA-EVLA baselines only, as it does not
correlate on ELVA-VLA baselines. 

To mitigate the above effects, 
we used the following modified approach to the gain calibration.
We first
applied hanning smoothing to the visibility data and discarded every
other channel, resulting in a 128-channel data set.
After applying the latest available corrections to the 
antenna positions and an initial excision of corrupted data, 
we  computed and applied an initial bandpass calibration to our
spectral line data to remove closure errors on 
VLA-EVLA baselines. EVLA-EVLA baselines were flagged during this step.
We then computed
a new frequency-averaged (channel~0) data set for use in calibrating
the frequency-independent complex
gains.  We excluded from this average the bottom $\sim$0.5~MHz portion of
band affected by aliasing (channels 1-79) as well as several edge
channels on the upper portion of the band (channels 121-128). 
After solving for the frequency-independent portion of the complex
gains using the newly computed channel~0 file, we computed
and applied an additional correction to the
bandpass, and then applied time-dependent frequency
shifts to the data to compensate for changes caused by
the Earth's motion during the course of the observations. 

Prior to imaging the line data, the $u$-$v$ data were
continuum-subtracted using a linear fit to the 
real and imaginary
components of the visibilities. Channels 21-32 and
59-120 were determined to be line-free and were used for these
fits. These channel ranges correspond to LSR velocities of
$-44.6$ to $-58.8$~\kms\ and $-93.6$ to $-172.1$~\kms, respectively. 
Although the spectral shape of the
aliased portion of the
continuum as measured toward our continuum calibrators was 
better approximated by a fourth order polynomial,
there was insufficient continuum signal in our line data to adequately
constrain such
high order fits. 

We imaged the VLA line data using the standard AIPS CLEAN deconvolution 
algorithm and produced data cubes using various 
weighting  schemes (Table~4). 
We also produced an image of the 21-cm continuum
emission in the X~Her field using the line-free portion
of the
band. The peak continuum flux density 
within the primary beam was
$\sim$0.35~Jy. We compared the measured flux densities of several
sources in the field with those from the NRAO VLA Sky Survey (Condon
et al. 1998) and found agreement to within $\pm$4\%.

We found no evidence for 21-cm continuum emission
associated with X~Her or its circumstellar envelope.
We detected a weak ($\sim$12~mJy), unresolved continuum source
offset
from the position of X~Her by \am{2}{4} and 
overlapping with the star's \HI\ emission. However, inspection of the Digitized
Sky Survey reveals that this most likely arises from
a pair of distant, interacting galaxies that are coincident with this
position. 

%
\begin{deluxetable}{lcccl}
\tabletypesize{\scriptsize}
\tablewidth{0pc}
\tablenum{3}
\tablecaption{VLA Calibration Sources}
\tablehead{
\colhead{Source} & \colhead{$\alpha$(J2000.0)} &
\colhead{$\delta$(J2000.0)} & \colhead{Flux Density (Jy)} & \colhead{Date}
}

\startdata
3C286$^{\rm a}$      & 13 31 08.2879 & +30 32 32.958 & 14.72$^{*}$   &
2007May 5 \& 10\\

1411+522$^{\rm b}$   &14 11 20.6477  &+52 12 09.141  &
22.27$\pm$0.20$^{*}$  & 2007 May 5
\\
...       & ...           & ...           & 22.07$\pm$0.14$^{*}$  &
2007 May 10\\

1625+415$^{\rm c}$  &16 25 57.6697  &+41 34 40.629  & 1.50$\pm$0.01 &
2007 May 5
\\
...       & ...           & ...           & 1.50$\pm$0.01 & 2007 May 10\\

\enddata

\tablecomments{Units of right ascension are hours, minutes, and
seconds, and units of declination are degrees, arcminutes, and
arcseconds. }
\tablenotetext{*}{Adopted flux density at 1420.3~MHz,
computed according to the VLA
Calibration Manual (Perley \& Taylor 2003).}
\tablenotetext{\dagger}{Quoted flux density is the mean from the
two observed frequencies; see \S~\ref{VLAobs}.}

\tablenotetext{a}{Primary flux calibrator.}
\tablenotetext{b}{Bandpass calibrator.}
\tablenotetext{c}{Phase calibrator.}

\end{deluxetable}

%
\begin{deluxetable}{lccccc}
\tabletypesize{\footnotesize}
\tablewidth{0pc}
\tablenum{4}
\tablecaption{Deconvolved Image Characteristics}
\tablehead{
\colhead{Image} & \colhead{{$\cal R$}} & \colhead{Taper} & 
\colhead{$\theta_{\rm FWHM}$} & \colhead{PA} & \colhead{rms} \\ 
\colhead{Descriptor} & \colhead{} & \colhead{(k$\lambda$,k$\lambda$)} & 
\colhead{(arcsec)} & \colhead{(degrees)} & \colhead{(mJy
beam$^{-1}$)} \\
\colhead{(1)} & \colhead{(2)} & \colhead{(3)} &
\colhead{(4)} & \colhead{(5)} & \colhead{(6)}  }

\startdata

Robust +1 & +1 & ... & $53''\times48''$ & $-37$ & 0.93-1.10\\

Natural & +5 & ...& $59''\times55''$ & $-42$ & 0.90-1.07\\

Tapered & +5 & 2,2 & $112''\times93''$ & $+47$ & 1.05-1.33 \\

Continuum & +1 & ... &   $52''\times48''$ & $-37$ & 0.26\\


\enddata

\tablecomments{
Explanation of columns: (1) image or data cube designation 
used in the text; (2) robustness
parameter used in image deconvolution (see Briggs 1995); 
(3) Gaussian taper applied in $u$ and
$v$ directions, expressed as
distance to 30\% point of Gaussian in units of kilolambda;
(4) dimensions of
synthesized beam; (5) position angle of synthesized beam (measured
east from north); (6) rms
noise per channel (1$\sigma$; line data) or in frequency-averaged
data (continuum).}

\end{deluxetable}

\section{Green Bank Telescope Observations\protect\label{GBTobs}}
To better characterize the extended emission distribution of
X~Her and its larger-scale 
environment,  we obtained \HI\ mapping observations of the 
X~Her field using
the 100-m Robert C. Byrd Green Bank Telescope (GBT) of the NRAO on
2008 February 21, 25, 26, and 27. 
The GBT spectrometer was employed with a 12.5~MHz bandwidth and 9-level
sampling. In the raw data, 
there were 16,384 spectral channels with a 0.7629~kHz (0.16~\kms)
channel width. 
In-band frequency switching was used with cycles of 0.8~Hz, alternating 
between frequency shifts of 0 and $-2.5$~MHz from the
center frequency of 1420.4058~MHz. This resulted in a usable LSR velocity
range from $-480$ to $+480$~\kms. Data were recorded in dual linear
polarizations. System temperatures during the run ranged from 15-18~K.
The spectral brightness temperature scale was determined from
injection of a noise diode signal at a rate of 0.4~Hz  and
was checked during each session with
observations of the line calibrator S6
(Williams 1973). 

We obtained data using two complementary mapping approaches. First,
to characterize the larger-scale
environment of X~Her, we mapped
a \ad{2}{2}$\times$\ad{2}{2} region centered on the position of the
star using
on-the-fly (OTF) mapping (see, e.g., Mangum et al. 2007). 
Second, to probe possible weak, spatially 
extended emission associated with 
the CSE of the star, we also
obtained deeper grid map observations of a $42'\times42'$ region
centered on X~Her. 
Both maps were obtained with \am{3}{5} (approximately Nyquist) 
sampling. 

For the OTF map, we scanned in right ascension and obtained two
complete passes over the entire region. Five seconds were
spent at each position during each pass, with a dump rate of
1.2~seconds. 

Reduction of the individual GBT spectra was performed using the
GBTIDL package. The total power for individual scans was computed using
the two signal spectra and the two reference spectra; these were first
combined to produce each calibrated spectrum and then folded to average the two
parts of the in-band frequency switched spectrum for increased 
signal-to-noise. Each processed spectrum was subsequently 
smoothed with a boxcar function with a kernel width of 5 channels and
decimated, resulting in 0.8~\kms\ channels. 
Lastly, a third order polynomial baseline was fitted to
the line-free portion of the spectrum and
subtracted. The channels used for the baseline fit were 1000 to 1350 and
1750 to 2100 in the decimated spectrum 
(corresponding to LSR velocity ranges of 118~\kms\ to 400~\kms\
and $-$123~\kms\ to $-$405~\kms, respectively).

The baseline-subtracted spectra were converted to FITS format using
the idlToSdfits program provided by G. Langston. The FITS data sets
were then loaded into AIPS for further processing and analysis.
After concatenating the spectra taken on different days, 
the individual spectra from the OTF and grid maps 
were respectively convolved and sampled onto regular grids 
using the SDGRD task, resulting in three-dimensional spectral line data
cubes.  For the gridding, a 
Bessel$*$Gaussian convolution function was used with
cell size of \am{3}{5}  (see Mangum et al. 2007).
The rms 
noise per pixel in the resulting maps was $\sim$19~mJy in the OTF data and
$\sim$3~mJy in the grid map data.

\section{Combining the VLA and GBT Data\protect\label{combo}}
Combining the GBT and VLA data allows a more complete
characterization of emission in the X~Her field over a wider variety of
spatial scales than either of these data sets alone. The GBT is well-suited
for filling in missing ``zero spacing'' information for VLA D
configuration observations since its diameter comfortably 
exceeds that of the minimum VLA baseline. Because of their larger area
coverage, we utilized the GBT OTF data
for this combination, together with the tapered VLA data (see Table~4).

As a first step for making a combined data set, 
the GBT OTF data were regridded to match the pixel scale of the deconvolved 
VLA data. The two data sets have different
spectral resolution ($\sim$0.8~\kms\ for the GBT data and
$\sim$1.2~\kms\  for the VLA
data). Because of the rather narrow line width of the star (see
Table~2), we did not
attempt to regrid the data in frequency, but instead manually paired each VLA
channel with the GBT channel closest in central velocity. This leads
to some frequency smearing, but this was
inconsequential for our purposes.

Next, the GBT channel images were multiplied by the VLA primary beam
pattern and blanked values were replaced with zeros. Finally the AIPS
task IMERG was used to create the combined channel images. IMERG
operates by fast Fourier transforming both data sets and combining them in
the Fourier domain. The single-dish image is normalized to the
deconvolved interferometer image using an overlap region in the
$u$-$v$ plane. This overlap annulus was selected to
be  35 to 100~m, or 0.166 to 0.476~k$\lambda$. The result was then
Fourier transformed back to the image plane. 
Results from the combined maps are discussed in \S~\ref{GBTresults}. 

\section{VLA Results\protect\label{vlaresults}}
\subsection{\HI\ Images of the X~Her CSE and Surrounding 
Field\protect\label{VLAHIprop}\protect\label{vlamom0}}
\begin{figure}
\vspace{-1.5cm}
\centering
\scalebox{0.8}{\rotatebox{0}{\includegraphics{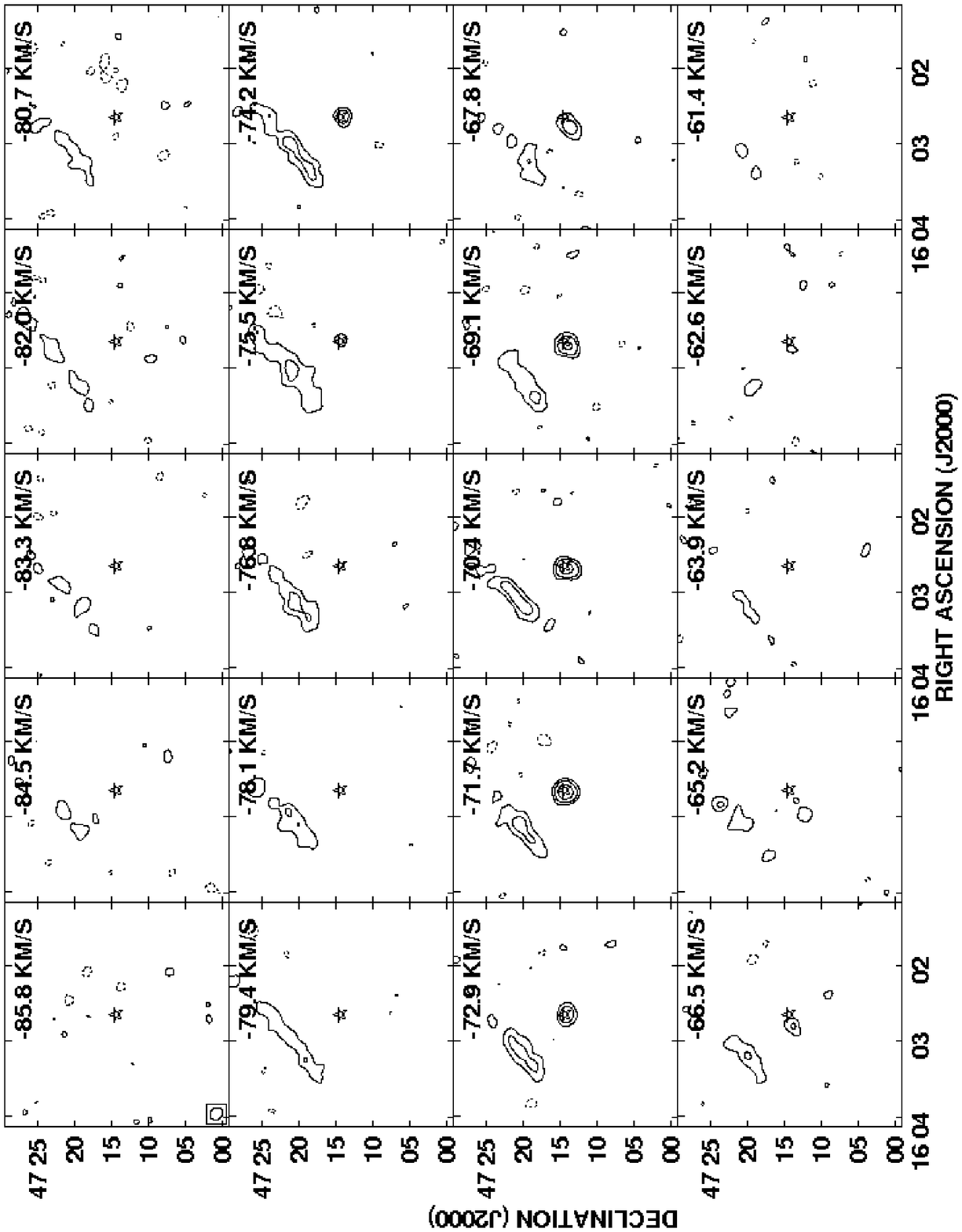}}}
\caption{\HI\ channel maps bracketing the LSR velocity of X~Her, derived
from the Tapered VLA data (see Table~4). 
A star symbol indicates the position of X~Her.
Contour levels are (-3,3,6,12)$\times$1.15~mJy beam$^{-1}$. The lowest
contour is $\sim3\sigma$. The range of velocities shown corresponds
approximately to the range of velocities over which CO(3-2) emission
was detected in the CSE of X~Her by Knapp et al. 1998. The
field-of-view shown is comparable to the VLA primary beam. The size of
the synthesized beam is indicated in the lower corner of the upper
left panel.
  }
\label{fig:vlacmaps}
\end{figure}

Figure~\ref{fig:vlacmaps} presents selected channel maps from a
tapered, naturally weighted VLA \HI\ data cube (see Table~4). 
The range of velocities
shown corresponds approximately to the range over which CO emission
has been previously detected in the CSE of X~Her
(e.g., Knapp et al. 1998). 
\HI\ emission is detected in all but one of these channels.
Toward the position of
X~Her, we detect \HI\ emission at $\ge 3\sigma$ significance in 10
contiguous channels, ranging in LSR
velocity from $-65.2$~\kms\ to $-76.8$~\kms, with the peak
emission occurring in the channel centered at $-71.7$~\kms. In the
same channels, as well as in several additional channels extended
toward higher negative velocities, we also detect extended emission
to the north and northeast of the star. 

\begin{figure}
\scalebox{0.4}{\rotatebox{-90}{\includegraphics{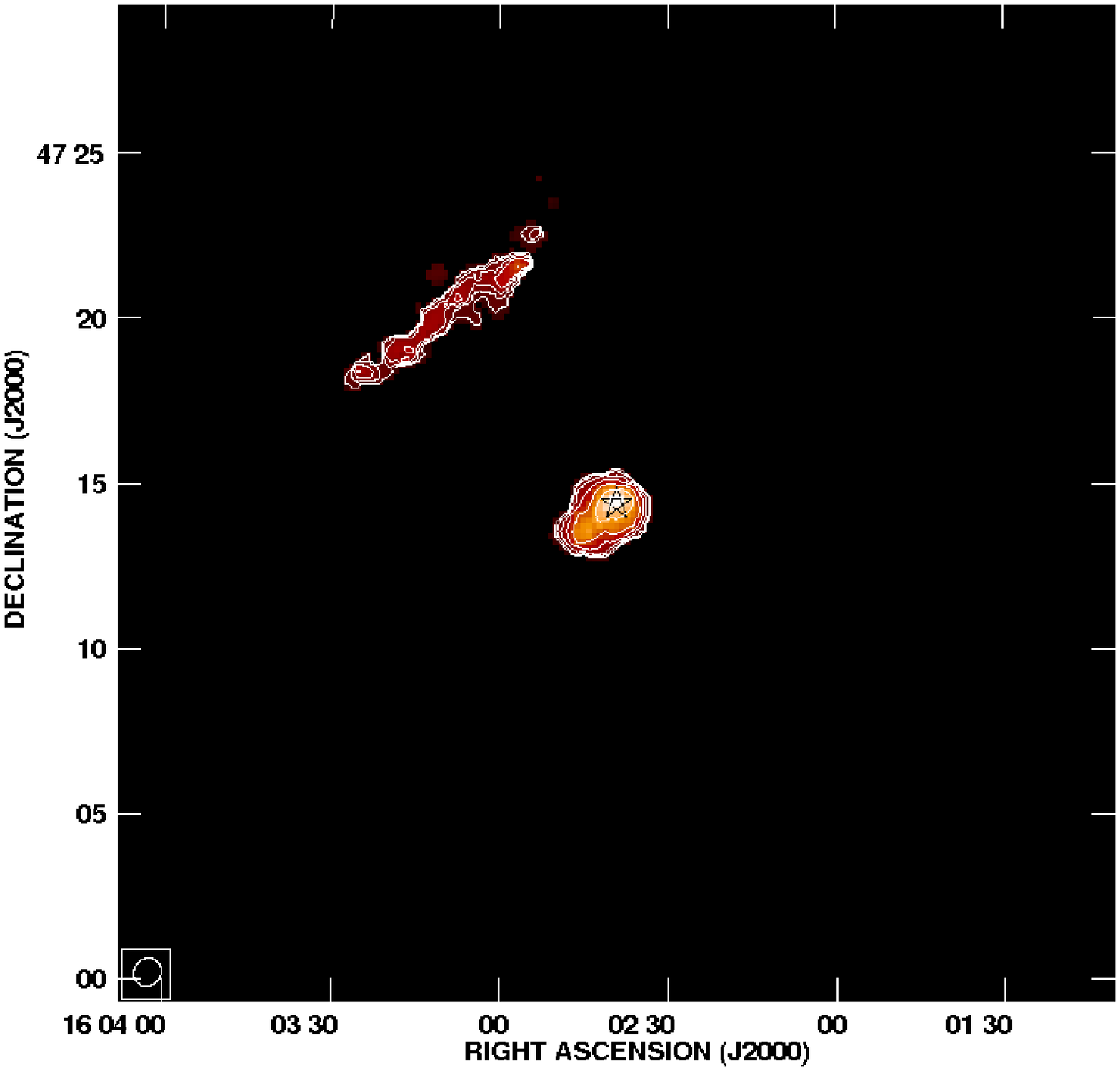}}}
\scalebox{0.4}{\rotatebox{-90}{\includegraphics{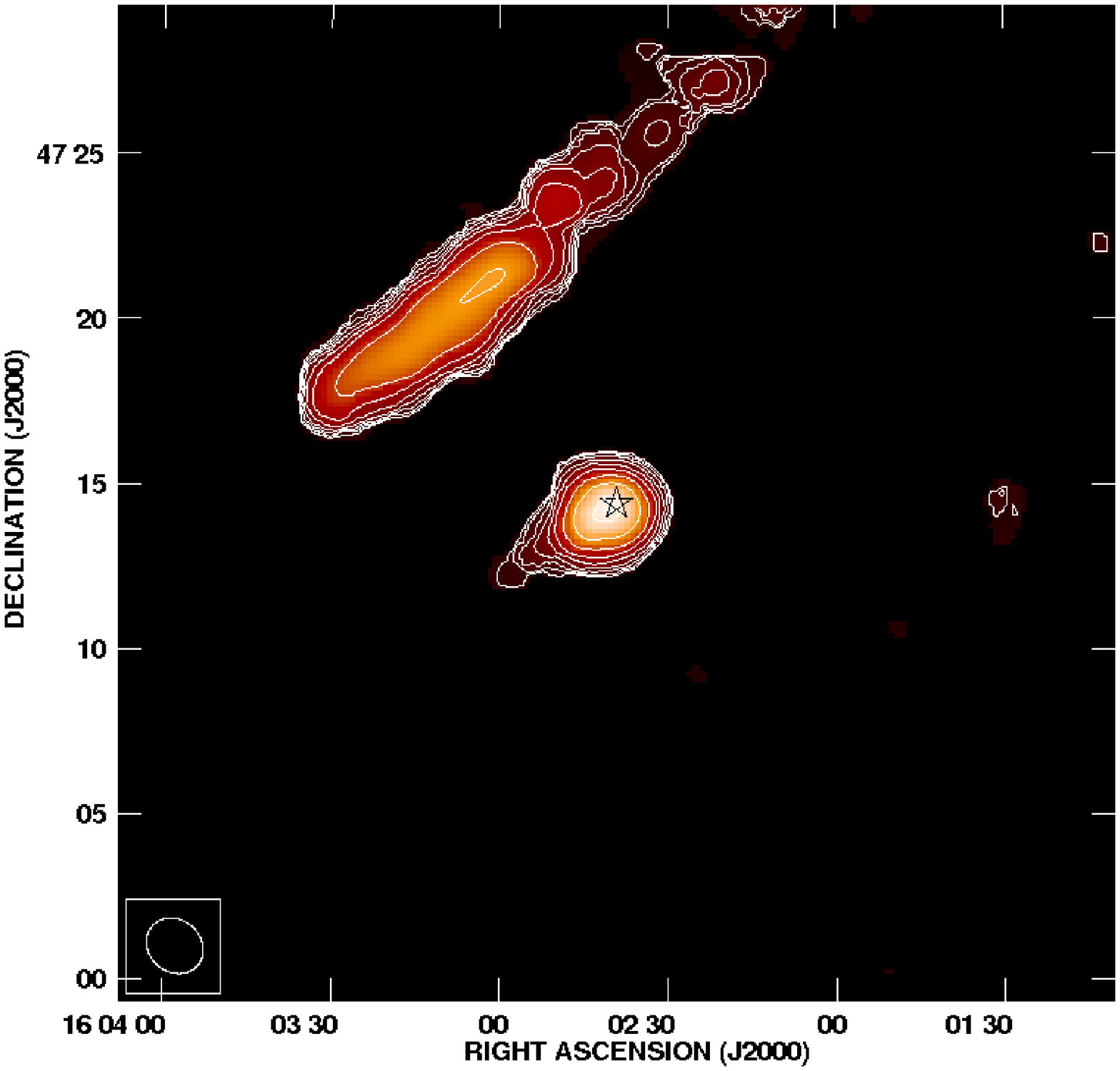}}}
\caption{\HI\ total intensity (zeroth moment)
maps of X~Her derived from the VLA Robust~+1
(left) and Tapered data (right) (see Tale~4). Star symbols
indicate the position of X~Her and the synthesized beam size is
indicated in the lower left corner of each panel.  Both
maps were constructed from emission spanning LSR velocities
$-75.5$ to $-65.2$~\kms. Intensity levels are 0 to 60 Jy beam$^{-1}$
m s$^{-1}$ (left) and 0 to 125 Jy beam$^{-1}$
m s$^{-1}$ (right).  Contour levels are
(1,1.4,2,...8)$\times$7~Jy beam$^{-1}$ (left) and
(1,1.4,2,...16)$\times$7~Jy beam$^{-1}$
m s$^{-1}$ (right). To minimize noise in the maps, data at a given point
were blanked if they did
not exceed a 2.5$\sigma$
threshold after smoothing by a factor of three
spatially and spectrally.  The stripe of emission to the
northeast is part of a much larger cloud (Cloud~I)
that does not appear to be directly
related to the CSE (see \S~\ref{GBTresults}). }
\label{fig:vlamom0}
\end{figure}

The distribution of detected \HI\ emission is further
illustrated in Fig.~\ref{fig:vlamom0}, 
where we show two \HI\ total
intensity (moment~0) maps of the region surrounding X~Her, derived
using two different versions of the VLA data cubes. 
Spectral channels spanning the velocity
range from $-65.2$~\kms\ to $-75.5$~\kms\ were used to produce these maps. 

In Fig.~\ref{fig:vlamom0}, 
we see that the emission to the
northeast of X~Her forms an elongated ridge that appears distinct from
the CSE in the VLA data. This
ridge is coincident with the  direction toward which Gardan et al. (2006)
reported evidence of a possible asymmetric outflow from
X~Her. However, with the VLA we do not detect any adjoining 
emission between the
ridge structure and the emission centered on
X~Her, suggesting that this material is not directly linked to its
CSE. Indeed,
as we describe below, our subsequent GBT mapping
observations have revealed that the ridge in fact appears to be a local
enhancement within a
larger \HI\ cloud that lies adjacent (in projection) to the star. 
We discuss this cloud and its possible origins 
further below. We focus the remainder of this section  on the
properties of the \HI\ envelope of X~Her as derived from our VLA
measurements. 

Fig.~\ref{fig:vlamom0} reveals
that the integrated emission
toward the position of X~Her is significantly extended relative to the
synthesized beam. Moreover, the emission exhibits a cometary or ``head-tail''
morphology, similar to what has been seen previously in
the \HI\ envelopes of RS~Cnc (Matthews \& Reid 2007) and Mira
(Matthews et al. 2008). The total extent of the X~Her
emission as measured from our tapered map (Fig.~\ref{fig:vlamom0}b) 
is $\sim6'$ (0.24~pc) at a limiting \HI\ 
column density of $\sim7.5\times10^{17}$ atoms cm$^{-2}$. The position
angle (PA) of the \HI\ ``tail'' at its outermost measured extent
measured from the tapered map is
$\sim121^{\circ}$.
However, measurements from the higher resolution map in
Fig.~\ref{fig:vlamom0}a yields a PA of $\sim132^{\circ}$ 
along the first $\sim3'$ of the tail, indicative  of  either a curvature or
a density asymmetry in the outermost tail material.

The total extent of the \HI\ tail measured by the VLA is approximately a
factor of two smaller than the size of the CSE previously derived from 
{\it IRAS} 60$\mu$m measurements ($R_{\rm outer}$=\am{6}{2}; Young et
  al. 1993). This suggests that the CSE may contain
even more extended gas to which the VLA is insensitive. This
possibility is consistent with our GBT mapping
measurements described below (\S~\ref{GBTstarspec}).

In the cases of Mira and RS~Cnc, the cometary 
morphologies of the \HI\ envelopes have been shown to arise 
from turbulent wakes that are formed as these
mass-losing stars barrel through the ISM (Matthews et al. 2008).
To evaluate whether a similar scenario can explain the morphology of
X~Her's CSE, we have computed
the components of the Galactic peculiar space motion of X~Her,
$(U,V,W)_{\rm pec}$, 
following the prescription of Johnson \& Soderblom (1987). For this
calculation we assume a heliocentric radial velocity of $-90.5$~\kms, a proper
motion in right ascension of $-68.49$~mas yr$^{-1}$, 
and a proper motion in declination of
64.65~mas yr$^{-1}$ (van Leeuwen 2007). 
Correction for the solar motion using the constants
of Dehnen \& Binney (1998) yields $(U,V,W)_{\rm
  pec}=(-61, -65, -36)$~\kms. Projecting back into an equatorial reference
frame yields 
$(V_{r},V_{\alpha},V_{\delta})_{\rm pec} = (-80, -42, 34)$~\kms. This
implies a space velocity for X~Her of 
$V_{\rm space}\approx 96$~\kms\ along a position angle of 
309$^{\circ}$. Thus both the position angle of the motion 
(roughly 180 degrees opposite of the position angle along which the
\HI\ tail extends)
and the relatively
high space velocity of X~Her, are consistent with the
interpretation of the \HI\ morphology as a gaseous wake trailing the star.
The velocity field of the tail material  (\S~\ref{vlamom1}) further
solidifies this interpretation. We note that because 
the dominant velocity component of
X~Her is radial, this results in an effective foreshortening of the tail
material from our viewing angle. Nonetheless, the tail material
contains a record of a significantly extended mass-loss history for
X~Her (see \S~\ref{tailage}).

\subsection{The \HI\ Velocity Field of X~Her\protect\label{vlamom1}}
\subsubsection{Evidence for an Interaction between X~Her and the
  Surrounding ISM}
In Fig.~\ref{fig:mom1} we present an \HI\ velocity field derived using
the same range of velocities as used to construct the total intensity
maps in
Fig.~\ref{fig:vlamom0}.  
The northeastern ridge of emission shows a
rather chaotic velocity pattern on these scales.  However,
across the \HI\ envelope of X~Her we see a clear, 
systematic velocity gradient of $\sim$6.5~\kms\ along the length of the 
emission (i.e., a projected gradient $\delta v\sim$35\kms~pc$^{-1}$). 
This is further highlighted in Fig.~\ref{fig:PV}, where we
plot a position-velocity (P-V) cut extracted along
the length of the \HI\ tail.  
The absolute value of the velocities decrease with
increasing distance from the star, which is consistent with
deceleration of swept-back wind material owing to its interaction with
the surrounding ISM (e.g., Raga
\& Cant\'o 2008; hereafter RC08). The same effect has been 
observed previously in the extended wake
of Mira (Matthews et al. 2008).

\subsubsection{A Possible Link between the Molecular Outflow and the 
Atomic Gas in the CSE?}
The CO(1-0) observations of Castro-Carrizo et al. (2010) 
show that the maximum velocity
gradient in the molecular outflow (which is confined to $r\lsim 10''$) 
is present
along a position angle of $\sim45^{\circ}$, with
blueshifted material to the southwest and
redshifted material to the northeast. A closer inspection of the \HI\ 
velocity field of X~Her (Fig.~\ref{fig:mom1}) reveals that 
near the stellar position, the isovelocity contours
are somewhat twisted and asymmetric relative to the bisector defined
by the trajectory
of the star's motion.  A P-V plot extracted from the ``robust +1''
version of the \HI\ data along
the same position angle as the CO outflow (Fig.~\ref{fig:ZPV}) shows
that despite our  limited spatial resolution, there is clear 
evidence for a velocity gradient of
$\sim8\pm2$~\kms\ with the same sense as seen in the CO data 
(i.e., the more blueshifted emission to the
southwest). The kinematics of the \HI\ material
are consistent with a possible
relationship to the molecular outflow, 
although the axis of symmetry appears to lie near $V_{\rm
LSR}\approx -72$~\kms\ in \HI\ compared with $V_{\rm LSR}\approx -73$~\kms\ in
CO, and the spatial resolution  of the
\HI\ data is too low to characterize the relationship between the
molecular and atomic gas in any detail. 

One possibile interpretation of Fig.~\ref{fig:ZPV} 
is that the \HI\ traces a continuation of 
the molecular outflow beyond the molecular dissociation radius. 
This would imply
that the dynamical age of the outflow is $\gsim$8000~yr 
(assuming $V_{\rm outflow}=5$~\kms). This is 
roughly an order of magnitude longer than
previous estimates based on CO data 
(cf. Kahane \& Jura 1996) and also exceeds by a factor of several or
more the typical dynamical 
ages of the bipolar flows observed in post-AGB stars 
(e.g., Huggins 2007).  Alternatively, the 
molecular outflow might be a more recent occurrence, with the
\HI\ sampling material from a previous (biconical or spherical) 
mass-loss episode.
\HI\ observations with higher spatial
and spectral resolution may help to provide additional insight.
We note that for the semi-regular
variable RS~Cnc, which also exhibits a bipolar molecular flow,
Libert et al. (2010b) found evidence for
an elongation in the \HI\ emission along the molecular outflow
direction, again consistent with the possibility of the \HI\ tracing a
continuation of the biconical molecular  outflow to larger scales.

Establishing the duration over which the mass-loss from
X~Her has been dominated by a bipolar rather than a spherical wind is
relevant to better understanding the CSE properties on large as well
as on small scales. For example,
Raga et al. (2008) have shown that the presence of a latitude-dependent
wind that is misaligned with respect to the direction of motion of a star
is expected to impact the large-scale structure of circumstellar
wakes.  Such an effect could plausibly
account for the apparent asymmetry of the
tail that we see at larger distances from X~Her (see
\S~\ref{VLAHIprop}). A prolonged bipolar outflow stage would also
impact calculations of the time-averaged mass-loss rate from the star,
since most  mass-loss rates are derived under
the assumption of spherically symmetric mass-loss (e.g., Knapp
et al. 1998). This would in turn impact the total predicted mass of
the \HI\ envelope and tail (see \S~\ref{tailage}).
 
\subsection{The \HI\ Spectrum and Total \HI\ Mass of X~Her Based on the VLA 
Data\protect\label{vlaHIspec}}
We have derived a spatially integrated \HI\ spectrum of X~Her from the
VLA data by 
summing the circumstellar emission 
in each channel image within an irregular ``blotch'' whose periphery was
defined by a 2.5$\sigma$ isophote.  Measurements were corrected
for the VLA primary beam. The resulting \HI\ spectrum, shown
in Fig.~\ref{fig:vlaHIspec}, exhibits a single-peaked, slightly
lopsided shape, similar in shape and breadth to the \HI\ spectra 
observed for a number of other AGB stars
(e.g., G\'erard \& Le~Bertre 2006; Matthews \& Reid 2007).  Modeling by
Gardan et al. (2006) showed that this type of observed asymmetry  
can be an additional hallmark of a mass-losing star's interaction with 
the ISM. For this reason, the spatially integrated \HI\ profile cannot 
be used to
reliably gauge the outflow speed(s) of X~Her, as its shape is
dominated by the effects of this interaction. 

Compared with the mean
position-switched NRT spectrum of X~Her presented by Gardan et
al. (2006), the peak flux density in the spatially-integrated 
VLA spectrum is $\sim$20\%
smaller, and the FWHM velocity width 
($\sim5.1\pm0.6$~\kms) is narrower. 
These differences result in part from our present exclusion of the emission
component to the northeast (which has a broader velocity
extent than the emission centered on the star; see
Fig.~\ref{fig:vlacmaps} and \S~\ref{GBTresults}). 
Based on the VLA data alone, we derive a 
velocity-integrated \HI\ flux density for X~Her of $\int Sd\nu =
0.24\pm0.01$~Jy~\kms, corresponding to a neutral hydrogen mass of
$M_{\rm HI}\approx 1.1\times10^{-3}~M_{\odot}$.
However, this value should be regarded as a lower limit, since 
our VLA maps appear to be missing some 
extended emission (see
\S~\ref{GBTstarspec}). 

\section{Results from the GBT Observations\protect\label{GBTresults}}
\subsection{Detection of a High-Velocity Cloud in the X~Her 
Field\protect\label{clouddetect}}
\subsubsection{\HI\ Properties of ``Cloud~I''}
Fig.~\ref{fig:gbtgridcmaps} presents selected \HI\ channel maps bracketing the 
velocity of X~Her, as derived from the GBT grid map data. 
These maps reveal that the elongated 
ridge of emission seen in our VLA maps (Fig.~\ref{fig:vlamom0})  
is part of a much larger-scale cloud of emission,
extending outside the VLA primary beam and
spreading well beyond the spatial scales of $\lsim15'$ to which the
VLA is sensitive. Following Gardan et al. (2006), we hereafter refer
to this entity as ``Cloud~I''.
At GBT resolution,
the peak emission from Cloud~I lies spatially to the northwest of
X~Her, and spectrally blueward of the stellar systemic
velocity. However, the dramatically improved surface brightness
sensitivity of the GBT reveals emission 
extending southward and passing through the
position of X~Her at all velocities 
where circumstellar \HI\ was
detected with the VLA (viz. $V_{\rm LSR}\sim -66$ to 
$-77$~\kms; Fig.~\ref{fig:vlacmaps}). 
Thus the CSE emission from
X~Her and the emission from Cloud~I are both spatially and spectrally
blended in the GBT data.

Cloud~I clearly extends beyond the northern edge of the 
GBT grid maps shown in Fig.~\ref{fig:gbtgridcmaps}. 
To more fully illustrate the size and large-scale morphology of
the cloud, we therefore show in 
Fig.~\ref{fig:OTFmom0} an \HI\ total intensity map of the region
derived from our GBT OTF mapping data. 
This latter map appears to 
contain the bulk of the cloud,
although it still does not encompass
its northernmost boundary or its eastern tip. Nonetheless, we see that 
the cloud has an elongated
shape, with its brightest
regions stretching along a position angle 
of $\sim310^{\circ}\pm5^{\circ}$. 
Note that the elongated isophotes defining the brightest emission 
in this
map lie to the north of the emission ridge seen in the VLA
map in Fig.~\ref{fig:vlamom0}, 
although both stretch along comparable position
angles.

To assess whether Cloud~I may have significant additional emission
lying outside of our map region, we have examined \HI\ maps extracted from
the Leiden/Argentine/Bonn (LAB) Survey (Kalberla et al. 2005)
over a several degree region surrounding the position and velocity of 
Cloud~I. The LAB data confirm that our GBT maps 
have covered the bulk of
Cloud~I. However, in the LAB maps we
find a second compact cloud (hereafter ``Cloud~II'') near $l^{II}=$\ad{77}{0},
$b^{II}$=\ad{48}{5} with a similar size and peak column density to
Cloud~I. Cloud~II appears
well-defined and distinct from Cloud~I 
at contour levels of $T_{\rm B}>0.3$~K, although it blends
with Cloud~I at lower surface brightness levels. This raises the
possibility that both of these clouds could be part of a larger
complex (see also Gardan et al. 2006). 
However, data of higher resolution and sensitivity will be
needed to evaluate this possibility and to 
more thoroughly characterize the properties of Cloud~II.

Based on the GBT OTF data for Cloud~I, we find that the peak
brightness temperature, $T_{\rm B,peak}=0.69$~K, 
occurs at a position of 
$\alpha_{\rm J2000}=16^{\rm h}01^{\rm m}13.865^{\rm s}$, 
$\delta_{\rm J2000}= 48^{\circ} 06' 44.76''$ and in the spectral 
channel centered
at $V_{\rm LSR}=-80.99$~\kms. The Jy to $T_{B}$ conversion factor
at the frequency of our observations is 
2.21 K Jy$^{-1}$. In the velocity-integrated total intensity map
presented in Fig.~\ref{fig:OTFmom0}, the peak column density 
is $N_{\rm HI}=2.8\times10^{19}$ cm$^{-2}$. 

To better illustrate the relationship between the emission seen in the
GBT maps and that detected in our VLA imaging, 
Figure~\ref{fig:combocmaps} presents
several sample channel maps showing the combined VLA+GBT data. 
These maps reveal that the \HI\ emission
from X~Her stands out clearly from the larger-scale cloud
emission despite their spatial overlap at several positions. 
These maps also illustrate the relationship
between the ridge of emission detected by the VLA to the northeast of
X~Her and the lower column density emission in the region.

\subsubsection{The \HI\ Spectrum of Cloud~I\protect\label{GBTspec}}
In Fig.~\ref{fig:gbtcloudspec} we show a spatially integrated
\HI\ spectrum of Cloud~I derived from the OTF data. Emission in each
channel was
summed within a fixed $84'\times63'$ 
rectangular aperture, centered at $\alpha_{2000}=16^{h}04^{m}11.5^{s}$,
$\delta_{2000}=+47^{\circ}47'$\as{29}{8}.
We see from Fig.~\ref{fig:gbtcloudspec} (upper panel) 
that the emission from Cloud~I is superposed atop a
blue wing extending from the
dominant Galactic \HI\ signal along the line-of-sight. The spatially
integrated line profile of Cloud~I
is well fitted by a Gaussian plus a linear background term,
resulting in the following global \HI\ line parameters:
peak flux density: $F_{\rm peak}=24.8\pm1.3$~Jy; line centroid:
$V_{\rm LSR}=-79.0\pm$0.1~\kms;
FWHM line width: $\Delta V_{\rm FWHM}=20.0\pm0.1$~\kms. The
line centroid thus lies blueward of the stellar systemic velocity of
X~Her and outside the velocity range over which statistically
significant emission was detected from the CSE of the star by the VLA.

Based on the Gaussian fit parameters, we derive an
integrated \HI\ line flux for Cloud~I 
$\int S_{\rm HI}d\nu=526$~Jy~\kms, implying 
that if it were at the same
distance as X~Her (140~pc), its 
total \HI\ mass would be $\gsim$2.4~$M_{\odot}$. This \HI\ mass is a lower
limit, since Cloud~I appears to extend slightly beyond our map region.
Such a quantity of gas is
too large to have been shed by X~Her alone (whose current mass
is estimated to be 1.9~$M_{\odot}$; Table~1), unless this
material was significantly 
augmented by gas swept from the ambient
ISM.  We discuss the
significance of this finding and further constraints on the nature of 
Cloud~I in \S~\ref{clouddiscussion}. 

\subsection{Recovery of Spatially
  Extended Emission from X~Her with the GBT and the Total \HI\ Mass of
  the CSE\protect\label{GBTstarspec}}
Despite the spatial confusion between Cloud~I and the circumstellar
emission from X~Her in the GBT data, 
we have been able to spectrally isolate the
circumstellar signal. Fig.~\ref{fig:minispectra} shows
a series of one-dimensional spectra, each extracted from a single pixel
across a 5$\times$5 pixel portion of the GBT grid map. An asterisk
indicates the position of the star. We first attempted to fit each
spectrum with a single Gaussian line plus a linear background
term. However, for several of the spectra, 
these fits left statistically significant
residuals. We therefore repeated those fits with the inclusion of a
second Gaussian component. The results are overplotted
on the respective spectra in
Fig.~\ref{fig:minispectra}. The values of the central velocity ($V$),
dispersion ($\sigma$), and amplitude ($A$) for each Gaussian component
fitted are also indicated.

In Fig.~\ref{fig:minispectra} we see that the spectra requiring a second
Gaussian component are clustered around the position of X~Her. In
all of these cases, one spectral component is narrower, of
lower amplitude, and (with one exception) 
systematically redshifted compared with a second,
broader line component. Further, the velocities and line widths of
these components match well with those derived for the circumstellar
emission from X~Her based on our VLA data (\S~\ref{vlaHIspec}). 
We therefore conclude
that we have unambiguously detected emission from the CSE of X~Her
with the GBT. 

We note that the  FWHM line widths and velocity centroids 
of the two line components fitted to the spectra in 
the vicinity of X~Her are comparable to those of the
``broad'' and the ``narrow'' line components identified by Gardan et
al. (2006) in their NRT spectra toward the star. However, 
our new analysis now reveals that the
broader of these two line components appears to be 
due primarily to Cloud~I rather
than a northeasterly, asymmetric outflow as originally proposed by
Gardan et al. Our present data thus provide no clear evidence for a distinct
broad \HI\ component associated with the CSE of X~Her.

In Fig.~\ref{fig:gbtstarspec} 
we show a global spectrum for X~Her that we derived
from the GBT data by summing the spectra from each position in
Fig.~\ref{fig:minispectra},  after
subtraction of the background and the emission attributed to Cloud~I. The VLA
spectrum from Fig.~\ref{fig:vlaHIspec} is overplotted for comparison. 
The peak in the VLA spectrum occurs at slightly
lower negative velocities compared with what is seen in the GBT
spectrum, but this difference is not statistically significant.
Although the FWHM line widths of the two spectra are comparable, the
integrated \HI\ line flux derived from the GBT data is $\int S
dv$=0.46~Jy~\kms, corresponding to 
$M_{\rm HI}\approx 2.1\times10^{-3}~M_{\odot}$, or roughly twice
the value derived from the VLA data. While this value is
subject to the uncertainties in the line decompositions
of the individual spectra, the significant increase compared with the
VLA measurement suggests
that the GBT has allowed us to recover extended emission
from within the CSE to which the VLA was not sensitive.
Moreover, we note that additional emission from the 
CSE may be present at several
of the positions along the lower and right-hand sides of
Fig.~\ref{fig:minispectra}; however, the emission at these
locations was too weak to permit an unambiguous decomposition of the line into
multiple Gaussian components, and we were thus unable to
ascertain whether it arises from the CSE, from Cloud~I, or a
combination of the two. 

While the coarse spatial resolution of our GBT maps make it impossible to
determine the overall size of the X~Her CSE to great precision, the
results presented in Fig.~\ref{fig:minispectra} clearly suggest that
the CSE is more extended than inferred from our VLA observations alone
(\S~\ref{VLAHIprop}). Indeed the \HI\ extent of the CSE now appears to
be marginally consistent with the radius $R_{\rm out}$=\am{6}{2}
derived from {\it IRAS} far-infrared (60$\mu$m) measurements by Young
et al. (1993). 
From the same {\it IRAS} measurements, Young et al. derived a
mass for the CSE of X~Her of 
$M_{\rm tot}\approx4\times10^{-3}~M_{\odot}$ (after
scaling to our currently adopted distance). If we correct our GBT \HI\
mass for the presence of He, our new total mass estimate of $M_{\rm
  tot}\approx3\times10^{-3}~M_{\odot}$ is in reasonable agreement with the
far infrared estimate.

\subsection{Evidence for Interaction of Cloud~I with the ISM}
\begin{figure}
\centering
\scalebox{0.7}{\rotatebox{0}{\includegraphics{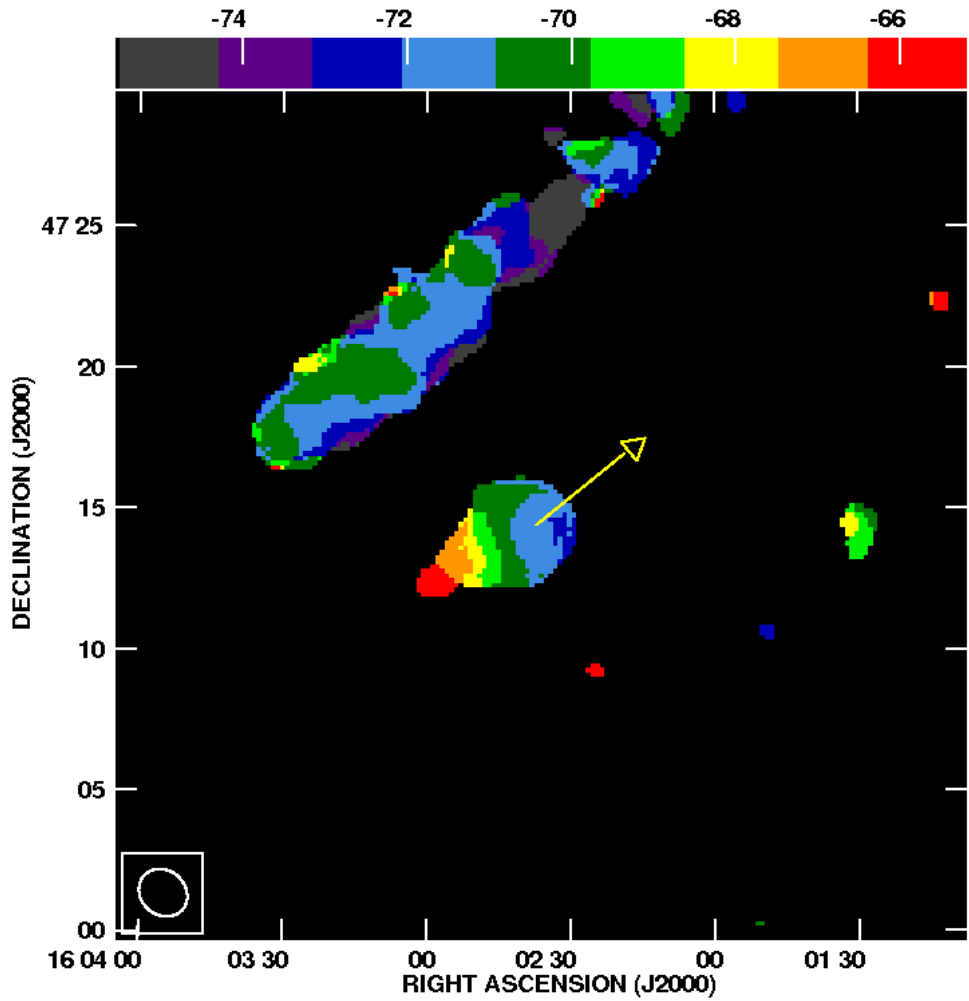}}}
\caption{\HI\ velocity field of X~Her, derived from the VLA Tapered
  data. The direction of X~Her's space motion  is indicated
  by an arrow. Colors indicate radial velocity (LSR frame) in \kms.
A systematic velocity gradient is seen along the emission
  trailing opposite to the direction of motion of X~Her. }
\label{fig:mom1}
\end{figure}

While isolating the emission from X~Her was the original goal of the
spectral decompositions of our GBT spectra, an
examination of the results as a function of position 
has revealed an interesting trend
(Fig.~\ref{fig:minispectra}). Near the position of X~Her, we find that
the centroids of the emission that we attribute to Cloud~I
(green lines) are redshifted relative to the mean velocity of the cloud
($\approx$80~\kms) and differ from X~Her by $\lsim$2~\kms. However, 
moving northward, the line
centroids of the Cloud~I emission become
systematically blueshifted, indicating the
presence of a systematic velocity gradient across the cloud. This trend is
further highlighted in Fig.~\ref{fig:gridmom1}, where we show an \HI\
velocity field derived from the GBT grid data. Emission in this map is
dominated by Cloud~I. (The signal-to-noise in the
OTF data was insufficient to construct a useful
first moment map).  Interestingly, the 
direction of the velocity gradient is similar
to that within the ``tail'' of
X~Her (Fig.~\ref{fig:mom1}). 

In Fig.~\ref{fig:OTFmom0} we saw that Cloud~I has an elongated
morphology that extends along a position angle that is similar
to the direction of motion of X~Her through the ISM (cf.
Fig.~\ref{fig:mom1}). This, coupled with the presence of a
velocity gradient along the direction of elongation, suggests that
the material in Cloud~I may be influenced by its motion
through the ISM. Furthermore, it raises the intriguing possibility
that Cloud~I and X~Her might share a
common space motion.

To investigate this further, we have examined some
P-V cuts extracted at several positions parallel to 
the long  axis of Cloud~I, 
spaced by $\sim7'$. The cut 
positions are indicated on Fig.~\ref{fig:OTFmom0}.  
We assume a position angle for the cloud of $309^{\circ}$ (the same
position angle as the space motion of X~Her; see \S~\ref{VLAHIprop}). 
The resulting P-V cuts (Fig.~\ref{fig:PVplots}) 
reveal several noteworthy kinematic features.
The lower panel of Fig.~\ref{fig:PVplots}
was extracted near the southwestern edge of the Cloud~I, and 
at this location we 
see that the emission comprising the cloud
(centered near $V_{\rm LSR}=-80$~\kms) 
exhibits an``S''-like shape, indicative of
higher negative velocities in the northwest
and lower negative velocities toward the southeast. This is 
consistent with the velocity
gradient seen in Fig.~\ref{fig:minispectra} \& Fig.~\ref{fig:gridmom1}. 

In the lower panel of Fig.~\ref{fig:PVplots}, Cloud~I 
appears to be well-separated from the primary
Galactic emission. However,
in the subsequent panels ($x\ge7'$), a ``bridge'' of emission becomes
visible, linking Cloud~I with the Galactic material. Furthermore, 
an additional ``streamer'' appears, stretching
along negative velocities to $V_{\rm LSR}\approx -65$~\kms.
In three of the panels ($x=7'$, $x=14'$, and $x=21'$), 
this lower streamer appears spatially
separated from Cloud~I, while in the top panel ($x=28'$) 
it blends smoothly with the
lower (southeastern) edge of the cloud. This lower streamer also corresponds in
position with a
compact feature near $V_{\rm LSR}\approx -$20~\kms\
(solid arrow on Fig.~\ref{fig:PVplots}). This compact feature is the most
prevalent at $x=14'$, but is visible in other P-V cuts as well.

The aforementioned features in the P-V plots along Cloud~I strongly
support our suggestion that Cloud~I has interacted significantly with
the ambient ISM. Indeed, we propose that the bridges, streamers, and
compact features described above are likely to
  be the result of
material that has been ram pressure stripped from Cloud~I as a
consequence of its space motion.
We discuss the
implications of this finding further in \S~\ref{clouddiscussion}.

\section{Constraints on the Molecular Gas and Dust Properties of 
Cloud~I}
\subsection{A Search for CO Emission\protect\label{COobs}}
To obtain additional constraints on the physical properties of Cloud~I, we have
used the Harvard-Smithsonian 1.2-m millimeter-wave telescope to
search for associated $^{12}$CO $J$=1-0 emission, which would indicate
the presence of a denser, molecular component to the cloud. These
observations were performed on 2008 March~13.

The observations were obtained using a 256
channel filter bank with 250~kHz (0.65~\kms) channel spacing, centered at a
velocity $V_{\rm LSR}=-70.0$~\kms.  The intensity scale was calibrated as
described in Dame et al. (2001) and is quoted in units of the main
beam brightness temperature. Frequency-switching
was employed, with an offset of 15~MHz (39~\kms) with
a cycle time of 2 seconds. 

The 1.2-m telescope has a FWHM
beamwidth of \am{8}{4} at 115.2712~GHz (Dame et al. 2001).
We obtained a 3$\times$3 grid map with one beamwidth 
separation between pointings, 
centered at  $(l,b)$=(\ad{75}{92},+\ad{47}{78}). The adopted center
is near the location of the peak \HI\ column density in Cloud~I as
inferred from our GBT observations (Fig.~\ref{fig:OTFmom0}). 
Total integration time was $\sim$5~minutes per pointing, resulting in an 
rms noise of $\sim$0.15~K per channel.

A spectrum derived from the sum of all 9 of our grid pointings is shown in
Fig.~\ref{fig:COspectrum}. A sixth order baseline has been fitted and
subtracted. The rms noise in this summed spectrum is 0.06~K.
No CO emission is evident over the
velocity range corresponding to the \HI\ emission from Cloud~I
(Figure~\ref{fig:gbtcloudspec}). A weak, narrow line is seen slightly
outside this window,
near $V_{\rm LSR}=-47$~\kms, but this feature is most likely spurious,
as no conjugate line is seen in the
frequency-shifted reference spectrum. (Such a line would appear as an apparent
``absorption'' feature near $V_{\rm LSR}=-8$~\kms\ in the
frequency-differenced spectrum shown in 
Fig.~\ref{fig:COspectrum}). 

To place an upper limit on the column density of molecular gas in the core of 
Cloud~I, we
assume a fiducial
CO line width of 5~\kms\ and a Galactic CO-to-H$_{2}$
conversion factor of $X = N_{H_{2}}/\int T_{\rm mb}({\rm CO})dV = 
1.8\times10^{20}$~cm$^{-2}$~K$^{-1}$~(\kms)$^{-1}$ (Dame
et al. 2001). 
This translates to a 3$\sigma$ upper limit on the
H$_{2}$ column density of
$N_{H_{2}}<1.3\times10^{20}$~cm$^{-2}$. 

The lack of CO emission toward Cloud~I is consistent with its modest
peak \HI\ column density, as well as the lack of
detected CO emission from other \HI\ clouds with similar properties
(Wakker et al. 1997; Dessauges-Zavadsky et al. 2007). As discussed
below, Cloud~I exhibits properties consistent with a 
high-velocity cloud, which are typically undetected in CO, with the
exception of a handful of examples that
appear to be associated 
with tracers of past or ongoing massive star formation (D\'esert et
al. 1990; Oka et al. 2008).

\subsection{A Search for Far-Infrared Emission from Cloud~I}
Using the {\it IRAS}-IRIS images (Miville-Desch\^enes \& Lagache
2005), obtained via the IRAS archive (http://irsa.ipac.caltech.edu),
we have searched for FIR (60$\mu$m)
emission associated with Cloud~I. While {\it IRAS} clearly detects 
patches of diffuse FIR emission overlapping in position with Cloud~I,
we find no obvious correspondence between either the morphology or the surface
brightness of the FIR emission and the column density of
the \HI\ emission in the region. In particular, the ridge of
emission exhibiting the highest \HI\ column density in
Fig.~\ref{fig:OTFmom0} does not have any obvious FIR
counterpart. Moreover, most of the \HI\ emission that could plausibly
be associated with FIR emission in the region around X~Her and Cloud~I
has $V_{\rm
  LSR}\approx$0~\kms\ (i.e., near the peak of the Galactic emission),
not near $V_{\rm LSR}\approx -$80~\kms. We conclude that based on
present data, there is no compelling evidence for a warm dust component
associated with Cloud~I.

\section{Discussion}

\subsection{The Age of X~Her's Tail and the Implications for 
its Mass-Loss History\protect\label{tailage}}
\begin{figure}
\centering
\scalebox{0.9}{\rotatebox{0}{\includegraphics{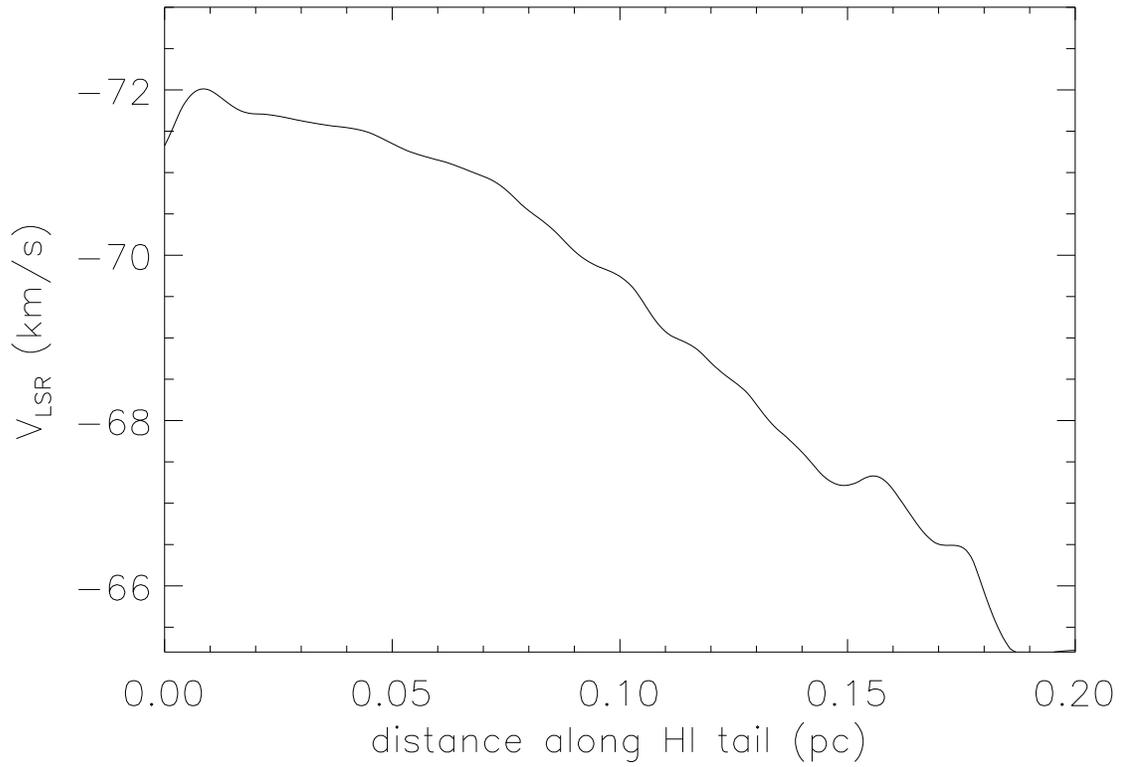}}}
\caption{\HI\ position-velocity cut along the ``tail'' of X~Her
  derived from VLA data. The cut was extracted along a position angle
  of 309$^{\circ}$. The plotted quantities are uncorrected for
  projection effects; distances are measured in the plane of the sky and
  velocities are along the line-of-sight. }
\label{fig:PV}
\end{figure}

\begin{figure}
\centering
\scalebox{0.65}{\rotatebox{-90}{\includegraphics{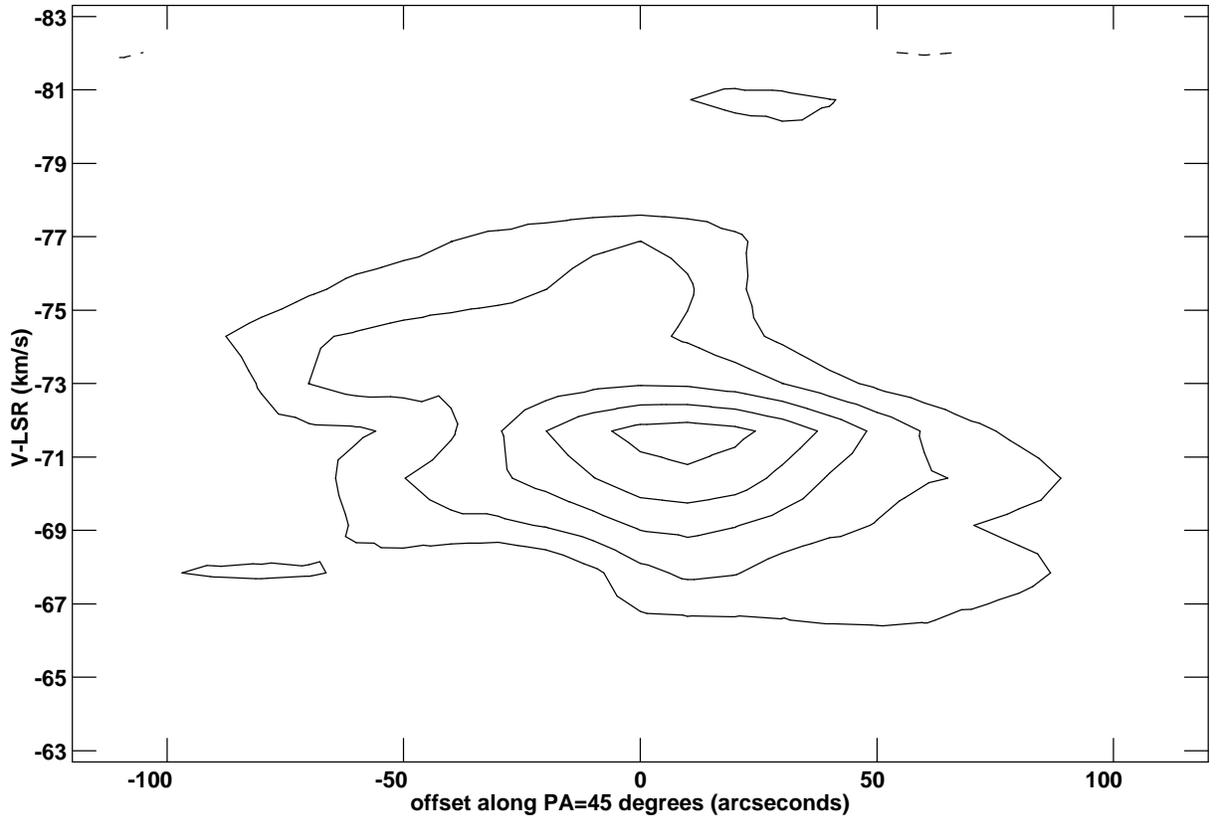}}}
\caption{\HI\ position-velocity plot for X~Her, 
extracted along a position angle
of 45$^{\circ}$ from the ``robust +1'' data (Table~4). The stellar
systemic velocity is $V_{\rm LSR}\approx-73$~\kms, and spatial offsets
along the $x$-axis 
increase to the northeast. A bipolar outflow with an extent of $\sim\pm10''$ 
has been seen along this direction in previous CO observations. 
The \HI\ emission extends over larger spatial
scales, but exhibits a velocity gradient of comparable sign and
magnitude to the CO emission. Contour levels are
($-2$,2,4,6,8,10,12)$\times1.0$~mJy beam$^{-1}$. }
\label{fig:ZPV}
\end{figure}

\begin{figure}
\centering
\scalebox{0.9}{\rotatebox{0}{\includegraphics{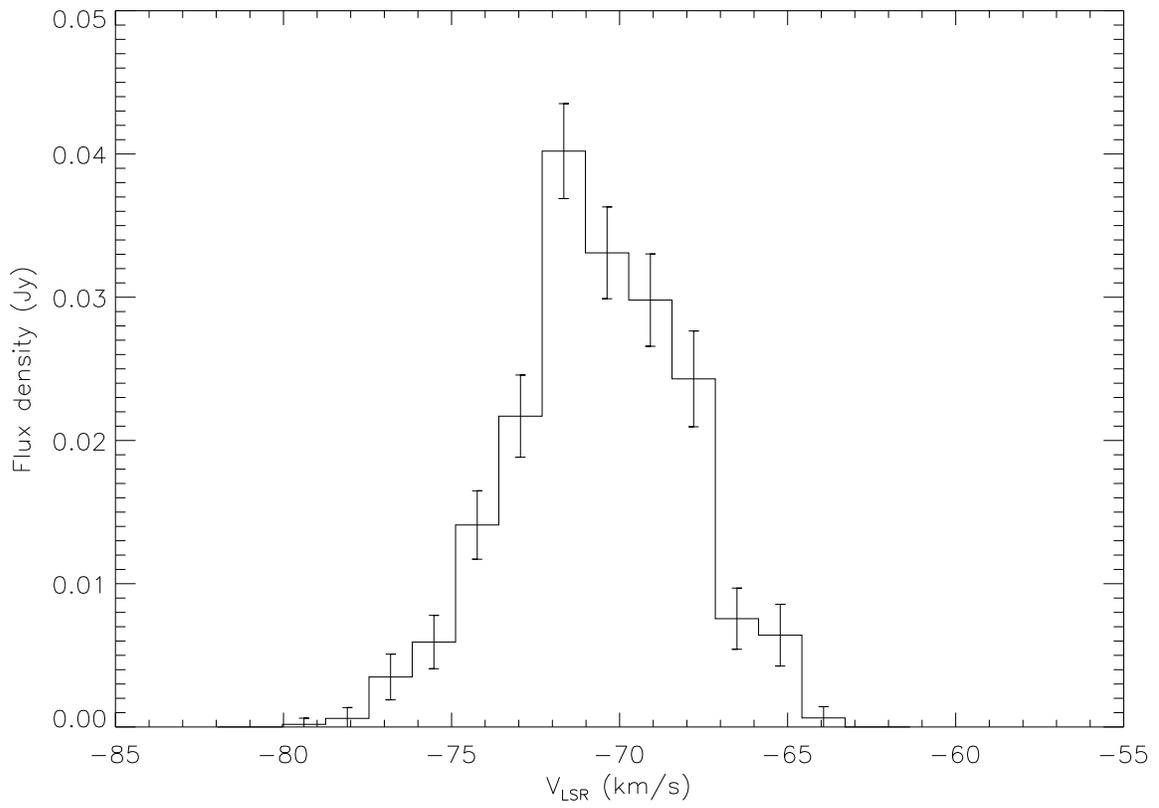}}}
\caption{Spatially integrated 
\HI\ spectrum of X~Her derived from the VLA data. }
\label{fig:vlaHIspec}
\end{figure}

\begin{figure}
\vspace{-1.0cm}
\scalebox{0.78}{\rotatebox{0}{\includegraphics{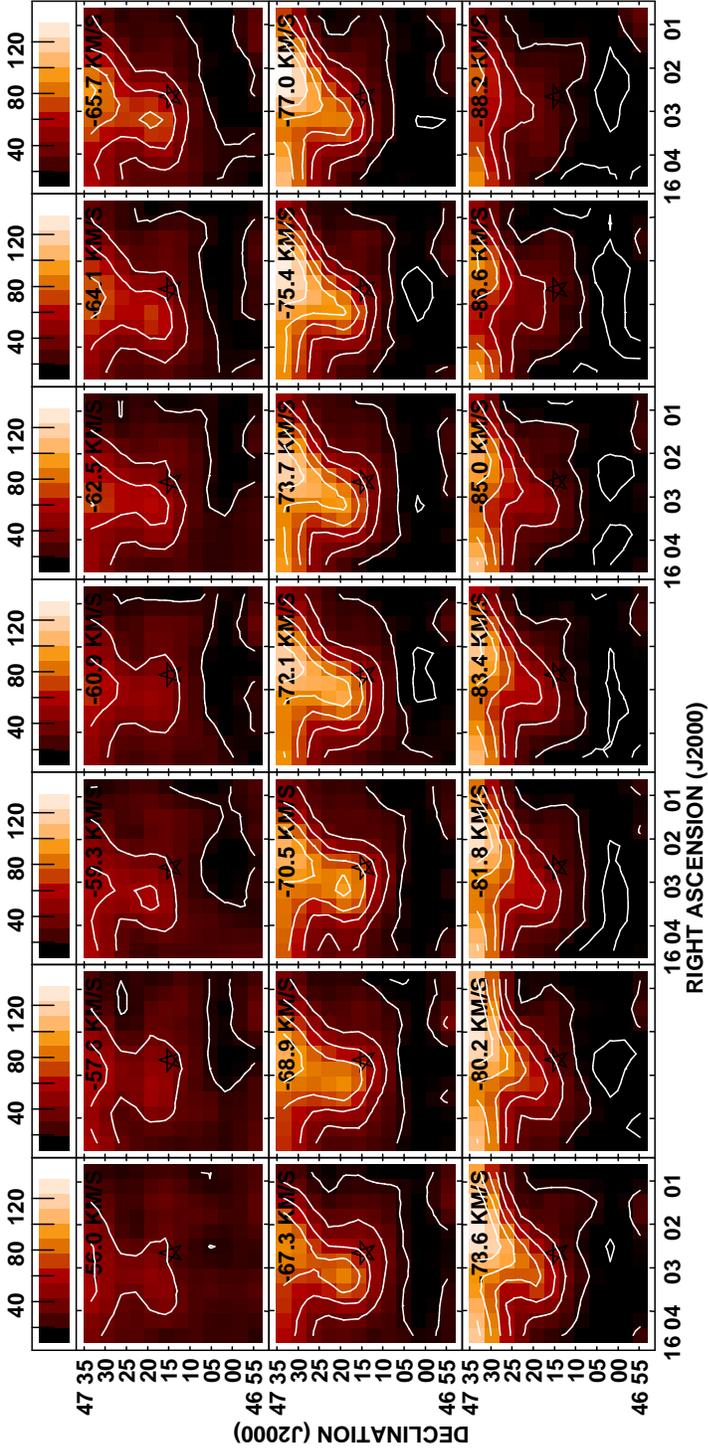}}}
\caption{\HI\ channel maps bracketing the velocity of X~Her, derived
from a GBT grid map of a $\sim40'\times40'$ region around the
star. Every third channel is shown within the plotted velocity range. 
Contour levels are ($-1$,1,3,6,9,12,15,18)$\times$6~mJy beam$^{-1}$ and
the intensity range is 0 to 140~mJy beam$^{-1}$. The lowest
contour is $\sim2\sigma$.
The GBT beam has FWHM $\sim9'$ and the pixels are
  \am{3}{5}. These maps reveal that the elongated 
ridge of emission seen in VLA maps (Fig.~\ref{fig:vlamom0}) 
is part of a larger cloud lying predominantly north of
X~Her, but which also extends southward
through the position of the star.  }
\label{fig:gbtgridcmaps}
\end{figure}

As described in Matthews et al. (2008), the 
kinematic information supplied by \HI\ observations of circumstellar
wakes is particularly valuable for estimating their ages,
and hence the overall duration of the AGB mass-loss
phase. Furthermore, the recent work by RC08 provides an
elegantly simple approach to computing the mass-loss duration, even for
the case where the deceleration of the wake material is non-uniform,
as appears to be the case for X~Her
(see Fig.~\ref{fig:PV}).

To estimate the age of X~Her's tail, 
we adopt a reference frame in which the star is
stationary, and the ISM appears to stream past it (see Fig.~2 of RC08). 
To perform this calculation, 
we have first deprojected the distances along the $x$-axis of
Fig.~\ref{fig:PV} and the measured radial velocities along the \HI\
tail ($y$-axis of Fig.~\ref{fig:PV}) 
into the X~Her rest frame using the space velocity vectors derived in
\S~\ref{VLAHIprop}. A subsequent fit to the resulting velocity versus
position curve yields the polynomial coefficients $a=0.09\pm0.01$ and
$d=1.5\pm0.1$~pc 
(see Eq.~5 of RC08), which in turn define an analytic law describing
$[V_{\rm space} - v(x)]$, where $V_{\rm space}$ is the space velocity
of the star (or equivalently, the streaming velocity of the ISM in the
stellar rest frame), and $v(x)$ is the average axial velocity along
the tail at a distance $x$ from the star. Finally, one may  express
the age of the tail as $t_{\rm tail}=\int^{x_{0}}_{0}[V_{\rm
    space} - v(x)]^{-1}dx$, where we have taken $x_{0}$=0.32~pc as the maximum
(deprojected) distance along the tail at 
which we were able to reliably measure the gas velocity. From
this approach, we find $t_{\rm tail}\approx1.2\times10^{5}$~yr. 
This new estimate for the mass-loss duration of X~Her exceeds by roughly
a factor of three the value
previously derived by Young et al. (1993) using infrared
measurements, and it approaches the expected interval between thermal
pulses for AGB stars (cf. Vassiliadis \& Wood 1993).

Sources of  uncertainty in our age estimate for the tail arise first,
from the uncertainty in its deprojected length, $x_{0}$, 
owing to errors on  the stellar parallax (5\%;
van Leeuwen 2007) and the proper motions ($\sim$1\%;
Perryman et al. 1997). A dominant source of 
uncertainty in the space velocity of X~Her arises from uncertainties in
the adopted solar constants (see
\S~\ref{VLAHIprop}), for which we estimate an error contribution
of $\sim\pm$5~\kms, or 5\%, to each of the equatorial components 
of the space velocity.  
The total formal error on $t_{\rm tail}$ is then 
$\sim$10\%. Lastly, the uncertainties in the coefficients of 
our polynomial fit contribute an additional
uncertainty in the age of $\sim$10\%. Combining the above values,  
we therefore 
estimate the global uncertainty in $t_{\rm tail}$ to be $\sim$20\%. We
note also that our age estimate depends implicitly on the
assumption that
the ambient ISM does not exhibit any local flows 
in the vicinity of X~Her.  

If we adopt a mass-loss rate for X~Her of
$\dot{M}=1.44\times10^{-7}~M_{\odot}$~yr$^{-1}$ (Table~1), our derived
age predicts that the total mass of the CSE of X~Her should be
$\sim0.017M_{\odot}$, or roughly 6 times greater than we have
inferred from our \HI\ measurements after correction for He
(\S~\ref{GBTstarspec}). One explanation for this discrepancy is that a significant
fraction of the CSE of X~Her comprises molecular rather than atomic
hydrogen.  However,
given the effective temperature of the star
($T_{\rm eff}\approx$ 3281~K), the models of Glassgold \& Huggins (1983)
predict that the wind should be predominantly atomic as it leaves the
star. This possibility therefore seems unlikely, unless there exists a
mechanism by which efficient H$_{2}$ formation can occur within the
tail (see also Ueta 2008). 

A second
possibility is that we have attributed too large a fraction of the
\HI\ emission in the vicinity of the star to Cloud~I. However, fully
reconciling our \HI\ data with the above mass prediction would require that 
{\em all} of the  emission detected by the GBT in the vicinity of
X~Her (i.e., within
panels 2, 3, and 4 of the lower three rows of
Fig.~\ref{fig:minispectra}) is associated with its CSE. This extreme
scenario seems unrealistic since Fig.~\ref{fig:combocmaps}
shows clear evidence for spatial overlap between stellar and cloud
emission. Further, we note that a total CSE mass of
$\sim0.017~M_{\odot}$ would require a highly unusual gas-to-dust ratio
to reconcile it with the FIR measurements of Young et al. (1993). 

A third conceivable explanation for the above discrepancy is that the
mass-loss rate from X~Her has not remained constant over the past
$\sim10^{5}$~yr. Indeed, the map shown in Fig.~\ref{fig:vlamom0}b
shows that
the \HI\ tail tapers to a significantly narrower shape with increasing
distance from the star, consistent with a
lower past mass-loss rate.   Furthermore, our 
age estimate for X~Her predicts that the star
should have undergone at least one thermal pulse, after which the
mass-loss rate is predicted to decrease by roughly an order of magnitude
(Vassiliadis \& Wood 1993).  Unfortunately,
direct evidence for time-variable mass-loss (e.g., in the form of
density variations in the tail material) is difficult to infer
from the present \HI\ data because of their 
limited spatial resolution.  Moreover, the
hydrodynamic simulations of Wareing et al. (2007) have shown that such
effects are likely to be  difficult to distinguish from
density variations in the tail resulting from turbulent effects. 

Finally, we should consider the overall uncertainty in the stellar mass-loss
rate, 
independent of whether this rate has been time-variable. Mass-loss
rates derived for X~Her by different workers based on SiO or CO data exhibit a
dispersion of $\sim$15\% (e.g., Table~2). However, the
uncertainties inherent in 
various model assumptions imply that a more realistic
estimate of the overall uncertainty in the mass-loss rates could be as
high as a factor of two
(see e.g., Olofsson et
al. 2002). This could conceivably account for part  of the discrepancy
between our measured and predicted values for the tail mass of X~Her. 
However, an additional complication is that all of the mass-loss rates in
Table~2 were derived under the assumption of
spherically symmetric mass-loss. To our knowledge, 
Kahane \& Jura (1996) are the
only authors to have derived a mass-loss rate for X~Her that takes
into account the bipolar nature of its molecular outflow
(\S~\ref{oldstuff}). When scaled
to our currently adopted distance, the mass-loss 
rate of Kahane \& Jura is comparable to or
higher than the values quoted in Table~2 
[i.e., $(1.4-3.4)\times10^{-7}M_{\odot}$ yr$^{-1}$, depending on the
adopted inclination], which could make the
``missing mass'' problem even more severe. 

\subsection{The Origin and Nature of Cloud~I and its Possible Relationship to 
X~Her\protect\label{clouddiscussion}}
We summarize in Table~5 some properties of Cloud~I as derived in the
preceding sections. 
It is noteworthy that
in terms of several of its global properties (\HI\ linewidth, peak \HI\
column density, and 
peak brightness temperature) Cloud~I shows a striking similarity
to the so-called ``compact high-velocity clouds''
(CHVCs; Braun \& Burton 1999). The CHVCs represent a distinct subset
of HVCs that are physically compact (FWHM$\le 2^{\circ}$) and 
sharply bounded in angular extent down to very low column
densities ($N_{\rm HI}\approx 1.5\times10^{18}$~cm$^{-2}$; de Heij et
al. 2002a). 
Based on a sample of 179 CHVCs, Putman
et al. (2002) reported mean values of $\Delta V_{\rm FWHM}$=35~\kms,
$N_{\rm HI}=1.4\times10^{19}$~cm$^{-2}$, and $T_{\rm peak}$=0.2~K,
respectively, comparable to values in Table~5. The integrated \HI\ flux of
Cloud~I is significantly larger than the mean value of 
19.9~Jy~\kms\ reported by Putman et al., suggesting that it may have a
smaller distance than most CHVCs,
although it still lies within the range of
reported values 
(see also de Heij et al. 2002c). Finally, like Cloud~I, 
nearly all CHVCs
discovered to date exhibit distinctly non-spherical shapes, which are
frequently consistent with ram pressure effects as the
clouds transverse the ambient medium (Br\"uns \& Westmeier 2004).

\begin{deluxetable}{lcc}
\tabletypesize{\scriptsize}
\tablewidth{0pc}
\tablenum{5}
\tablecaption{Properties of Cloud~I}
\tablehead{\colhead{Parameter} & \colhead{Value} & \colhead{Comment}}
\startdata

$\alpha$ (J2000.0) & 16 01 13.86 & $a$ \\

$\delta$ (J2000.0) & 48 06 44.76 & $a$ \\

$l$ & \ad{75}{81} & $a$ \\

$b$ & \ad{+47}{84} & $a$ \\

$T_{B}$ (peak) & 0.69~K& $b$ \\

$N_{\rm HI}$ (peak) & 2.8$\times10^{19}$ cm$^{-2}$ & $c$ \\

$\int S_{\rm HI}d\nu$ & 526~Jy~\kms & $d$\\

$M_{\rm HI}$  & $>1.22\times10^{-4} d^{2}_{*}~M_{\odot}$ & $d$,$e$\\

$V_{\rm LSR}$ & $-79.0\pm0.1$~\kms\ & $d$ \\

$\Delta V_{\rm FWHM}$ & 20$\pm0.1$~\kms\ & $d$ \\

Angular Size & $\sim2^{\circ}$ &... \\

$T_{k}$ & $\le$7750~K & $f$ \\

$N_{H_{2}}$ & $<1.3\times10^{20}$~cm$^{-2}$ & $g$ \\

\enddata

\tablecomments{Tabulated parameters are uncorrected for possible emission
  extending beyond our survey region. }

\tablenotetext{a}{Coordinates correspond to the location of the
peak detected flux 
density within a single spectral channel.}

\tablenotetext{b}{Peak in the spectral channel centered at $V_{\rm
    LSR}=-80.99$~\kms}.

\tablenotetext{c}{Peak in the velocity-integrated map under the
  assumption the gas is optically thin.}

\tablenotetext{d}{Based on Gaussian fit to the spatially integrated
  line profile.}

\tablenotetext{e}{$d_{*}$ is the distance in units of parsecs.}

\tablenotetext{f}{Derived using the relation
  $\sigma_{v}=\left(kT_{k}/m\right)^{0.5}$, where $k$ is the Boltzmann
constant, $m$ is the mass of hydrogen, and 
$\sigma_{v}$ is the gas velocity dispersion, taken to be 
$\approx$8~\kms. This value for $T_{k}$ is an upper limit
since it does not include corrections for 
possible non-thermal contributions to the line
  width.}

\tablenotetext{g}{3$\sigma$ upper limit.}

\end{deluxetable}

Cloud~I is not included in existing
CHVC catalogues (de Heij et al. 2002b; Putman et al. 2002), 
as it lies just outside the
velocity cutoff searched in previous surveys ($|V_{\rm LSR}|>$90~\kms;
Putman et al. 2002).
Adoption of such cutoffs in HVC
surveys is necessary to minimize
contamination and confusion by unrelated Galactic disk
emission. However, as pointed out by de Heij et al. (2002c), any reasonable
model for CHVCs predicts that examples should also be found at less
extreme velocities. The properties of Cloud~I together with 
its high Galactic latitude
suggest that it is one such example.  For this reason, we resist terming 
Cloud~I instead as an ``intermediate velocity cloud'' (IVC; see Wakker
2004), particularly since some evidence suggests that 
many IVCs are linked with energetic phenomena such as
supernovae and may
form a population largely distinct from the HVCs (D\'esert et
al. 1990; Kerton et al. 2006). 

Uncovering the nature of Cloud~I is complicated by the uncertainty in
its distance. A plausible lower limit for the cloud would be that
it is at the same distance as X~Her ($\sim$140~pc). If we assume that
the cloud consists primarily of neutral atomic material
(\S~\ref{COobs}), then at this
short distance its total mass would be only
$M_{\rm tot}\sim 3.4 M_{\odot}$ (where we have used a factor of 1.4
to correct the measured \HI\ flux density for the mass of He). 
As we noted in \S~\ref{GBTspec}, this quantity of material excludes the
possibility of an origin tied directly to the mass-loss from X~Her
unless the material was augmented significantly from debris swept
from the ambient ISM. Moreover, such a scenario would require a
differential acceleration between Cloud~I and X~Her such that Cloud~I
now has a larger blueshifted velocity compared with X~Her.  

Based on its morphology and kinematics,
Cloud~I is almost certainly not virialized; indeed, an implausibly
large distance of
$\sim$2~Mpc would be required in order that its virial mass, $M_{\rm
  vir}=\sigma^{2}r/G$, equal its
\HI\ mass, assuming a local velocity dispersion
$\sigma\sim$8~\kms\ (see Fig.~\ref{fig:minispectra}). Here
$r$ is the radius of the cloud and $G$ is the gravitational
constant. 

One further constraint on the distance to Cloud~I can be estimated from
our present data by assuming
that Cloud~I has interacted with material near its present location 
that follows the ``normal''
Galactic rotation pattern (see also Lockman et al. 2008). 
From Fig.~\ref{fig:PVplots}, we see
signatures of perturbations in material at velocities as high as $V_{\rm
  LSR}\lsim -10$~\kms. Following 
Reid et al. (2009),  the kinematic distance to material at
this position and velocity is $\sim4.98^{+0.71}_{-0.82}$~kpc. 
Unfortunately, kinematic
distance determinations toward the direction of Cloud~I are
highly uncertain, and 
changes of only a few \kms\ in the assumed LSR velocity can
translate to a
significant change in the implied kinematic distance. For example,
taking instead $V_{\rm
  LSR}=0$~\kms\ (the mean Galactic velocity toward this direction) 
implies either a ``near'' distance of
0.430$^{+4.21}_{-0.43}$~kpc (consistent with X~Her to within the large
uncertainties) or a ``far'' distance of
3.66$^{+0.98}_{-0.98}$~kpc.
We conclude that while kinematic arguments  suggest a rough upper limit to the
distance to Cloud~I of $\sim$5~kpc, they cannot yet exclude the possibility of 
a comparable
distance for Cloud~I and X~Her.  Future stellar absorption line
measurements toward Cloud~I 
and/or a search for associated diffuse H$\alpha$ emission may help to
provide additional distance constraints. 

As with the ``near'' distance for Cloud~I discussed above,
assumption of the ``far'' distance would seem to
lead to some puzzling inconsistencies. For example, at a distance of
5~kpc, Cloud~I would have an \HI\ mass of
$\sim3.1\times10^{3}~M_{\odot}$ and a linear extent of $\sim$170~pc,
which are both consistent with upper limits derived for CHVCs under
the assumption that they reside in the Galactic halo (Westmeier et al.
2005). On
the other hand, this would require that X~Her and Cloud~I are related
only by a chance superposition along the line-of-sight---a
circumstance with a rather low statistical probability.
Putman et al. (2002) found that the covering fraction for CHVCs is only
$\sim$1\%. Furthermore, toward a given Galactic longitude, a
typical dispersion in the observed distribution of velocities is
$\sim$100~\kms\ (see Fig.~20 of Putman et al.). Based on these statistics, we
estimate the probability of a chance superposition of a CHVC along the
line-of-sight to X~Her, having a peak velocity within $\pm$10~\kms\ of the
star, is approximately 0.2\%. 
While this estimate ignores possible clustering of CHVCs and does not
account for incompleteness of present surveys, finding an
apparent space motion for the cloud along the same direction as the
star's space motion makes the coincidence even more remarkable.

One possible explanation for a
CHVC in the vicinity of one or more evolved, mass-losing stars is if both have
been stripped from a globular cluster. For example, van Loon et al. (2009)
have identified a CHVC-like cloud in the vicinity of the globular
Pal~4 and have shown that a link between the two is 
plausible. However, the proximity of X~Her to the Sun and the Galactic
Plane  suggest that it is likely to be a disk object and therefore 
unlikely to have originated in a 
globular cluster. 

\section{Summary}
We have presented \HI\ 21-cm line observations of the CSE of the 
semi-regular variable star X~Her. 
X~Her was previously detected in \HI\ by Gardan et al. (2006), but our
new observations have allowed us to characterize the morphology and
kinematics of the emission in significantly greater detail. We find
that the \HI\ emission comprises an
extended, cometary-shaped wake 
that results from the motion of this mass-losing star
through the ISM. We estimate the total deprojected extent of the
extended CSE
to be $\ge$0.3~pc. Analogous \HI\ wakes have now been found associated with
other AGB stars including RS~Cnc (Matthews \& Reid 2007) and Mira
(Matthews et al. 2008), and evidence is accumulating that they 
may be quite common around mass-losing
stars, particularly those with high space velocities.

We detect a radial velocity gradient along the \HI\ tail of X~Her of
$\sim$6.5~\kms\ ($\delta v\sim$35~\kms~pc$^{-1}$), indicating a
deceleration of the circumstellar gas owing to its interaction with the ambient
ISM. Based on the observed deceleration, we estimate an age for the
tail of $\sim 1.2\pm0.3\times10^{5}$~years, implying a 
mass-loss history for the star several times longer than the previous
empirical estimate. However, the total \HI\ mass that we measure for the tail
($M_{\rm HI}\approx2.1\times10^{-3}~M_{\odot}$) is significantly
smaller than predicted given this age and the current 
mass-loss rate of the star derived from CO observations. 
One explanation is that the mass-loss
rate of the star has not been constant and has been increasing with
time. Another is that a significant portion of the tail material is in
the form of molecular hydrogen. 

Previous CO observations of X~Her have established the existence of a
bipolar outflow extending to $\sim\pm10''$ along a position angle of
45$^{\circ}$ (e.g., Castro-Carrizo et al. 2010). Along this position
angle, we have detected a
velocity gradient in the \HI\ emission 
similar to the one seen in CO, suggesting that the bipolar outflow may
extend to at least $\sim\pm1'$ from the star and have a dynamical age 
much larger than previously estimated ($\gsim$8000~yr). An alternative
is that
the \HI\ may be tracing an earlier mass-loss episode. 

X~Her lies (in projection) along the periphery of a more massive and
extended \HI\ cloud (Cloud~I), whose radial
velocity overlaps with that of the CSE of X~Her. Emission from
this cloud can account for the apparent asymmetric mass-loss from X~Her
previously reported by Gardan et al. (2006). We have characterized
the \HI\ properties of Cloud~I in detail and find them to be
similar to those of CHVCs. Using $^{12}$CO observations, we have also
placed an upper limit on the molecular gas content of the cloud of
$N_{H_{2}}<1.3\times10^{20}$~cm$^{-2}$. In our \HI\ observations,
we see evidence of a velocity gradient
along Cloud~I, suggesting that like X~Her, its morphology and
kinematics are affected by its motion through the ambient ISM. 
Both the high mass of Cloud~I ($M_{\rm HI}\gsim 2.4~M_{\odot}$) and
the blueshift of its mean velocity relative to X~Her appear to rule out
a direct circumstellar origin. However, the probability of a mere chance
superposition in position, velocity, and direction of 
space motion between Cloud~I and X~Her
is extremely low. Therefore a kinematic association between
the two objects cannot yet be excluded and awaits more robust distance
constraints for the cloud.

\acknowledgements

We thank 
J. Lockman for valuable guidance on the GBT observing setup and
data processing, as well as the entire
Green Bank staff for their hospitality and 
assistance with the
GBT observations. We also thank M. Reid and E. Greisen for useful 
discussions and the anonymous referee for suggestions that helped to
improve the manuscript.
MCJ was supported through a National Science
Foundation Research Experience for Undergraduates 
grant to MIT Haystack Observatory. 
The observations presented here were part of NRAO programs AM887 and GM75.
This research has made use of the SIMBAD database,
operated at CDS, Strasbourg, France and the 
NASA/IPAC Infrared Science Archive, 
which is operated by the Jet Propulsion Laboratory, California
Institute of Technology, under contract with the National Aeronautics
and Space Administration.

\clearpage

\begin{figure}
\hspace{1.0cm}
\scalebox{0.7}{\rotatebox{0}{\includegraphics{f8.ps}}}
\caption{\HI\ total intensity map of the region around X~Her, derived
  from the GBT OTF data, illustrating the morphology of the 
emission associated with Cloud~I.
The data were summed over the velocity range
$-95.4$ to $-56.0$~\kms. 
The intensity scale has units of Jy m s$^{-1}$ and is shown
with a logarithmic transfer function. Contour levels are
($-2$[absent],2,3,4...11)$\times$528~Jy m s$^{-1}$. The lowest contour
level is $\sim10\sigma$. The yellow ellipses indicate the
approximate location and extent of the \HI\ envelope of X~Her detected with the
VLA, and the ridge of Cloud~I emission
detected by the VLA, respectively  
(see Fig.~\ref{fig:vlamom0}). The yellow hatch marks indicate the
angles and locations along which the position-velocity cuts shown
in Fig.~\ref{fig:PVplots} were extracted.}
\label{fig:OTFmom0}
\end{figure}

\begin{figure}
\centering
\scalebox{0.7}{\rotatebox{-90}{\includegraphics{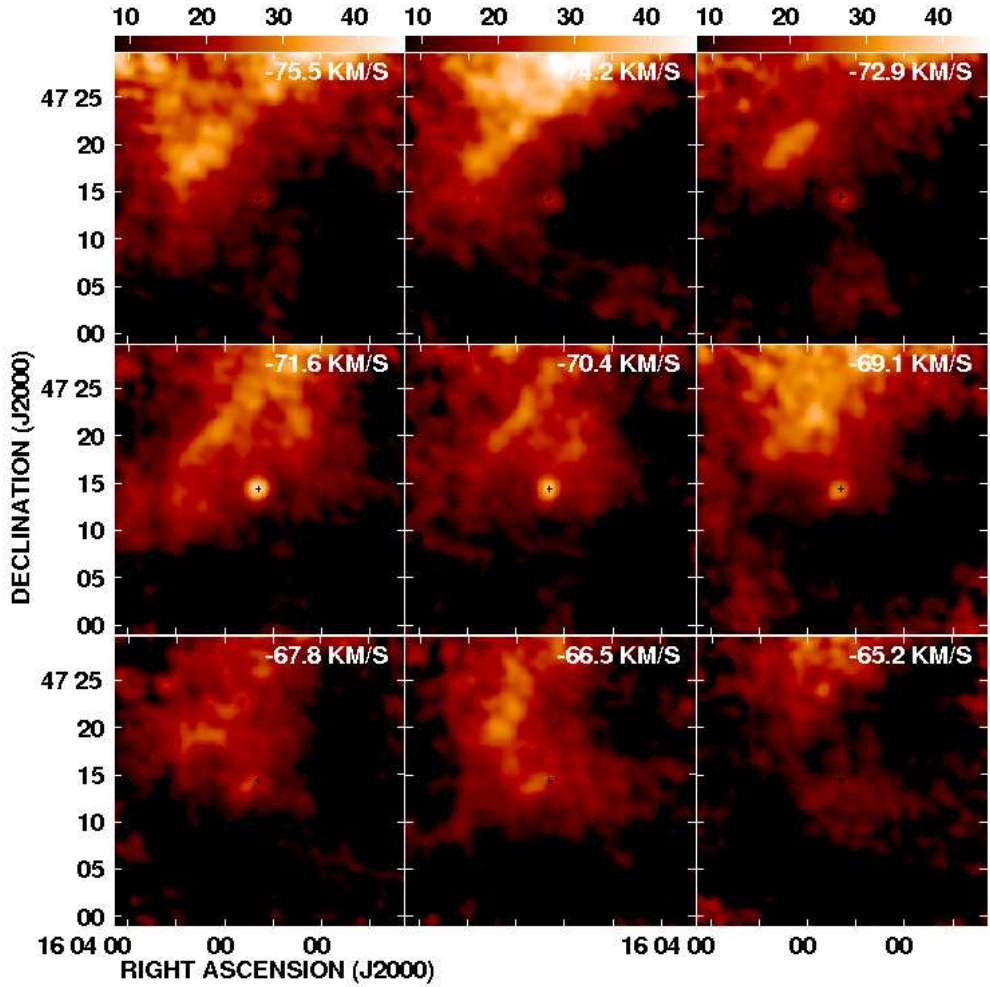}}}
\caption{Selected channel maps showing the combined GBT+VLA data
  across a $\sim30'$ field-of-view. The
  maps have been corrected for the attenuation of the VLA primary beam. The
  intensity range is 8 to 45.7~mJy beam$^{-1}$. A black cross
  indicates the position of X~Her. LSR velocities are indicated on
  each panel. }
\label{fig:combocmaps}
\end{figure}

\begin{figure}
\scalebox{0.7}{\rotatebox{0}{\includegraphics{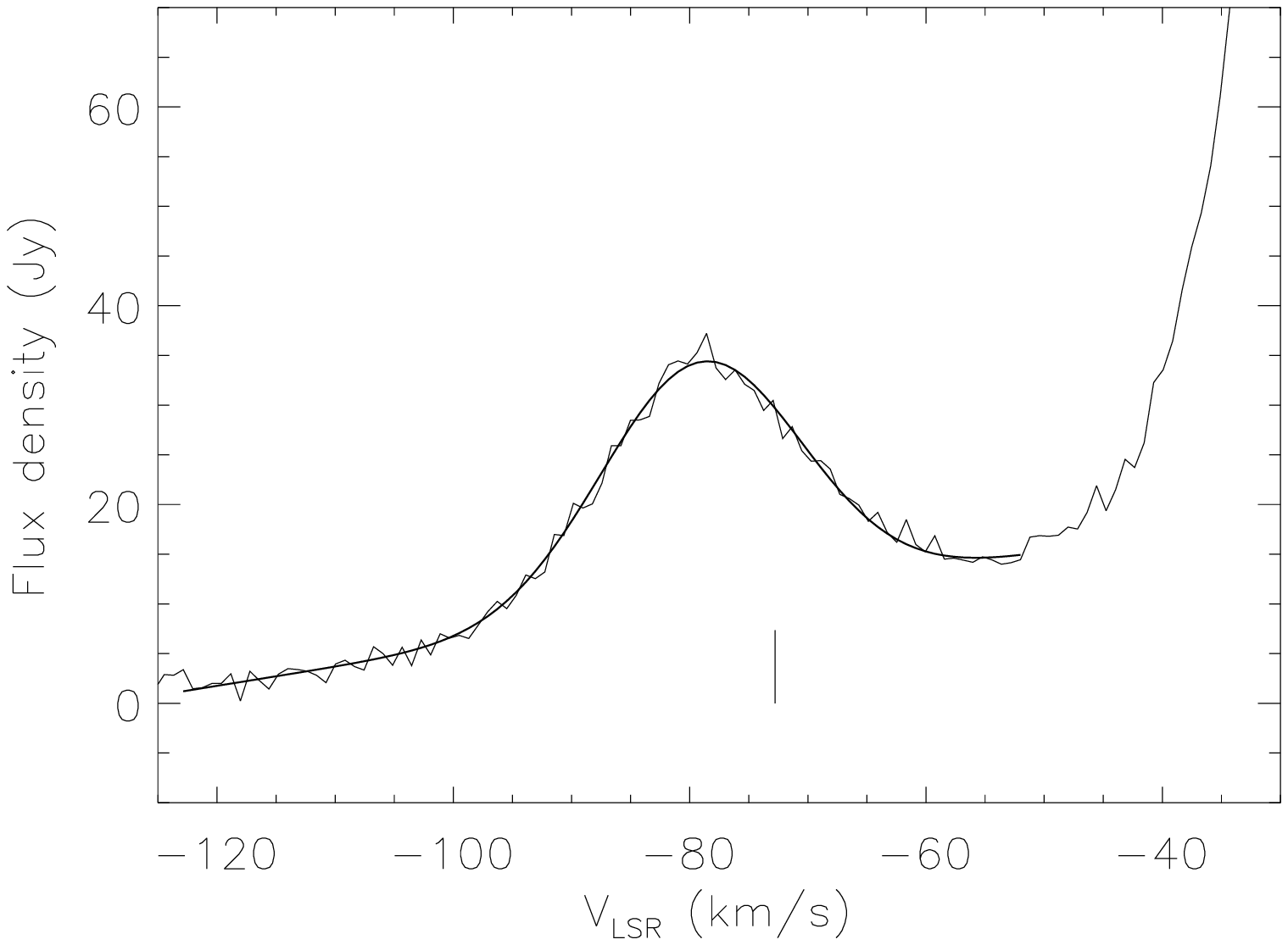}}}
\scalebox{0.7}{\rotatebox{0}{\includegraphics{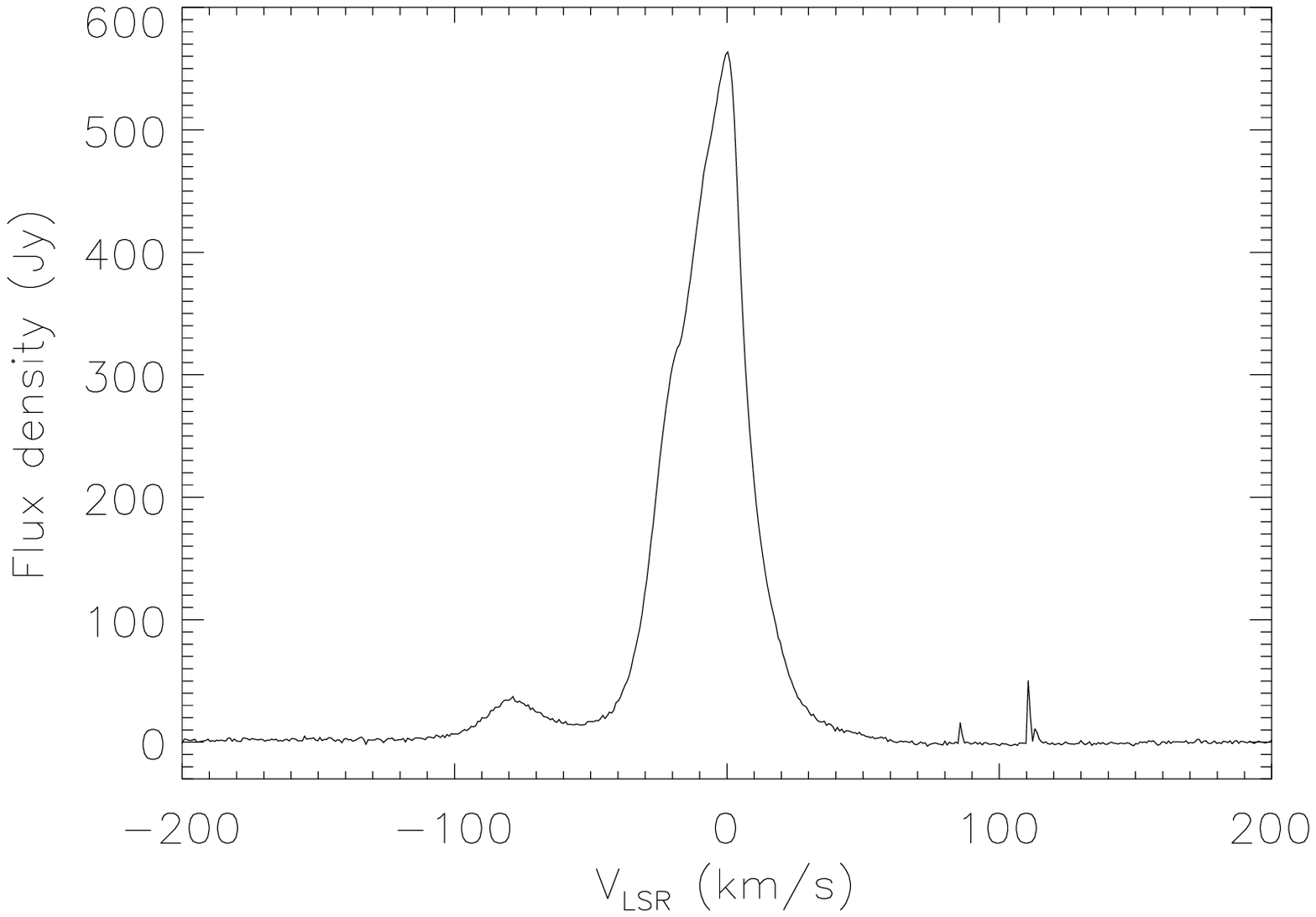}}}
\caption{Spatially integrated 
\HI\ spectra of Cloud~I derived from GBT OTF data (see Text). 
The upper panel shows a zoom of the spectral region around
Cloud~I. The systemic velocity of X~Her is indicated by a vertical
line. A fit to the Cloud~I line profile with a Gaussian plus linear
background is overplotted as a thick line. The
lower panel shows the same data over a broader velocity
range. The dominant Galactic emission toward this direction peaks near
$V_{\rm LSR}\approx$0~\kms. The spikes near 85 and 110~\kms\ are due
to RFI.  }
\label{fig:gbtcloudspec}
\end{figure}

\begin{figure}
\vspace{-0.5in}
\scalebox{0.75}{\rotatebox{0}{\includegraphics{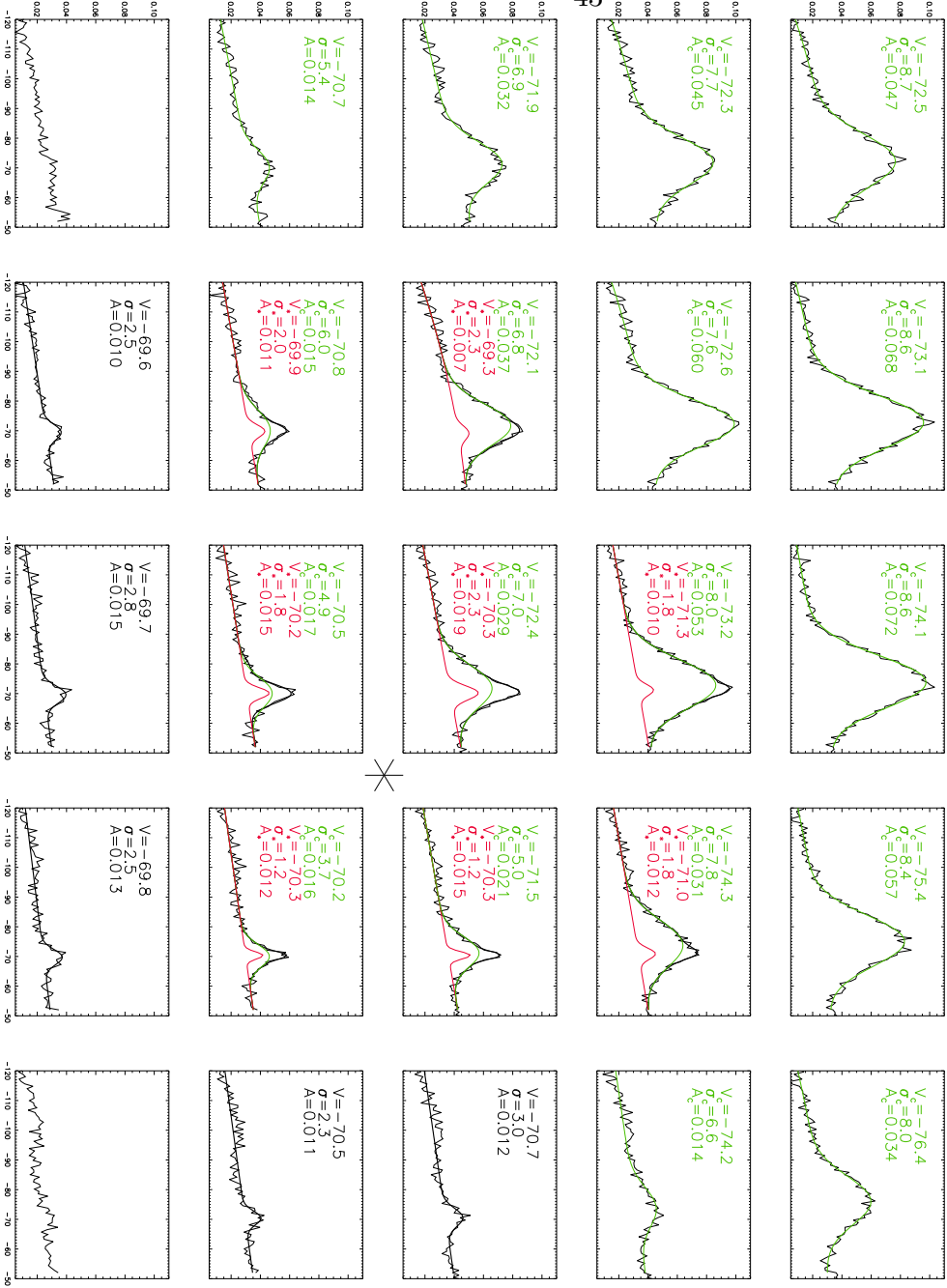}}}
\caption{\HI\ spectra extracted through each \am{3}{5} 
pixel of a 5$\times$5 pixel
  region of the GBT grid map. Axes are LSR velocity in \kms\ and flux
  density in Jy. The
  position of X~Her is indicated by a large asterisk. Overplotted are
  the results of Gaussian+linear background fits to each
  spectrum. The component of the 
emission attributed to Cloud~I is indicated by a thin green line and that
  attributed to the CSE of X~Her is indicated by a thin red line. In
  cases where two components were fitted, the sum is indicated by a
  thick black line. Finally, for cases where signal-to-noise was
  insufficient to distinguish between a one- and a two-component fit, 
a single Gaussian+background fit is overplotted in black. 
Fit parameters (velocity centroid in \kms, dispersion in
  \kms, and amplitude in Jy) for each fitted component 
are indicated on each panel in a coordinating font color.}
\label{fig:minispectra}
\end{figure}

\begin{figure}
\scalebox{0.9}{\rotatebox{0}{\includegraphics{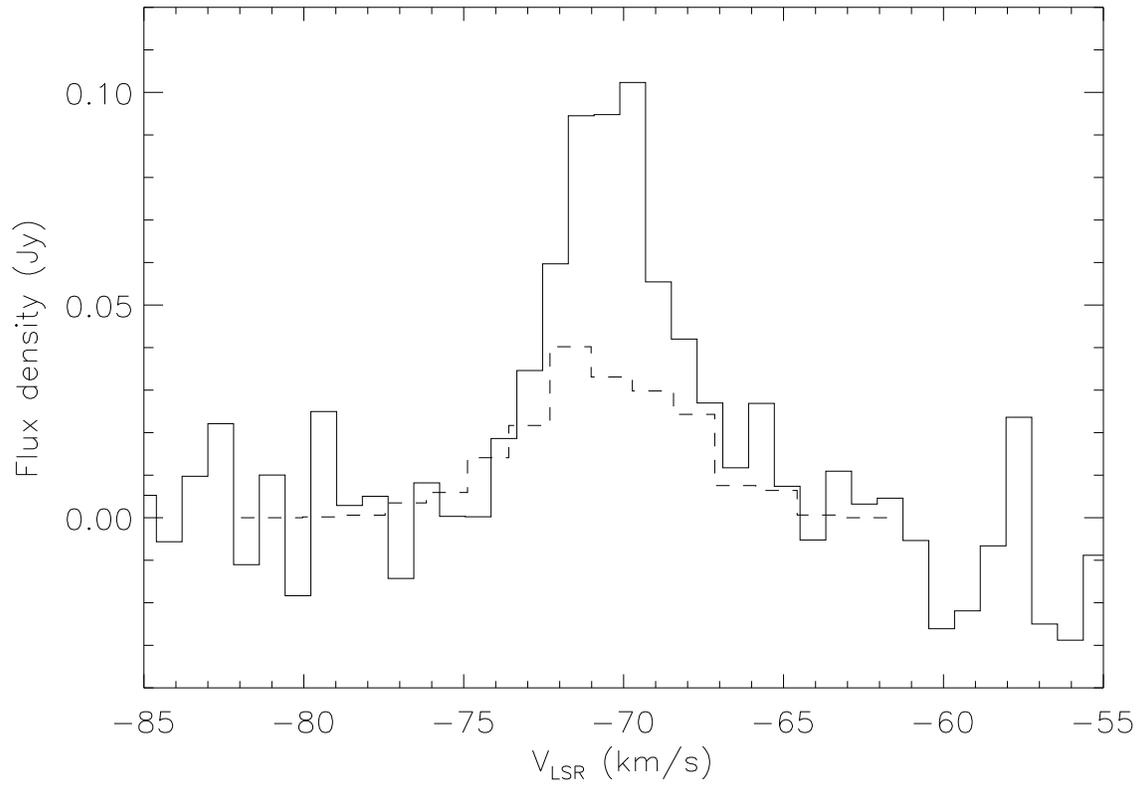}}}
\caption{Global
\HI\ spectrum of X~Her derived from the GBT grid data (solid line). 
The spectrum was derived by summing the seven spectra from
Fig.~\ref{fig:minispectra} where circumstellar emission could be
unambiguously identified. The emission contributions from Cloud~I plus
a linear background term were subtracted from each spectrum 
before summation.
The VLA spectrum from Fig.~\ref{fig:vlaHIspec}
is overplotted as a dashed line. }
\label{fig:gbtstarspec}
\end{figure}

\begin{figure}
\scalebox{0.9}{\rotatebox{0}{\includegraphics{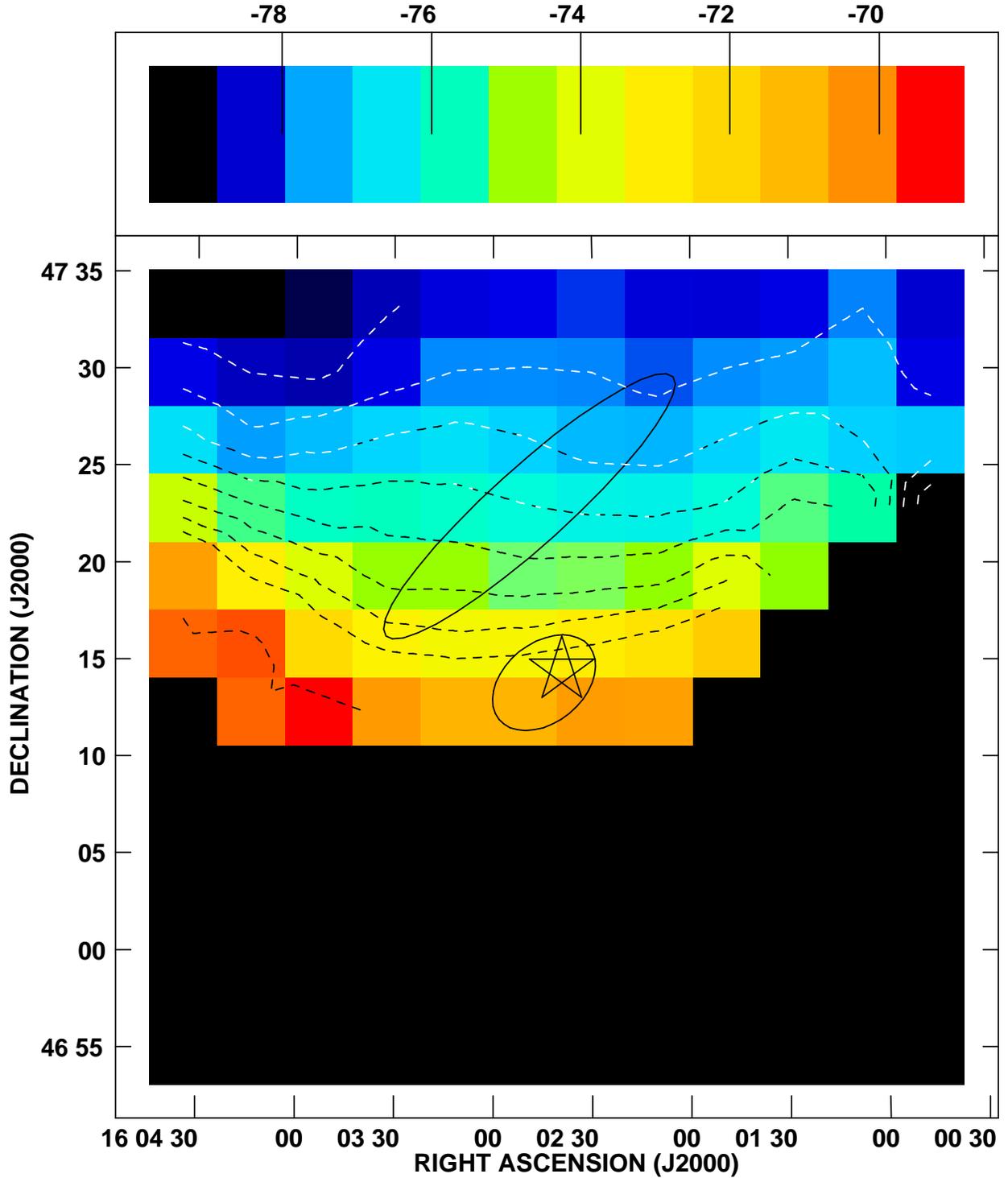}}}
\caption{\HI\ velocity field derived from the GBT grid data. The color
  scale is in units of 
LSR velocity in \kms. Emission in this map is dominated by Cloud~I. 
A systematic velocity gradient is present
across the map. The dark ellipses indicate the approximate location
  and extent of
the \HI\ envelope of X~Her detected with the VLA, and the ridge of
  Cloud~I emission detected with the VLA, respectively 
(see Fig.~\ref{fig:vlamom0}).}
\label{fig:gridmom1}
\end{figure}

\begin{figure}
\vspace{-10.0cm}
\hspace{-1.0cm}
\scalebox{1.8}{\rotatebox{-90}{\includegraphics{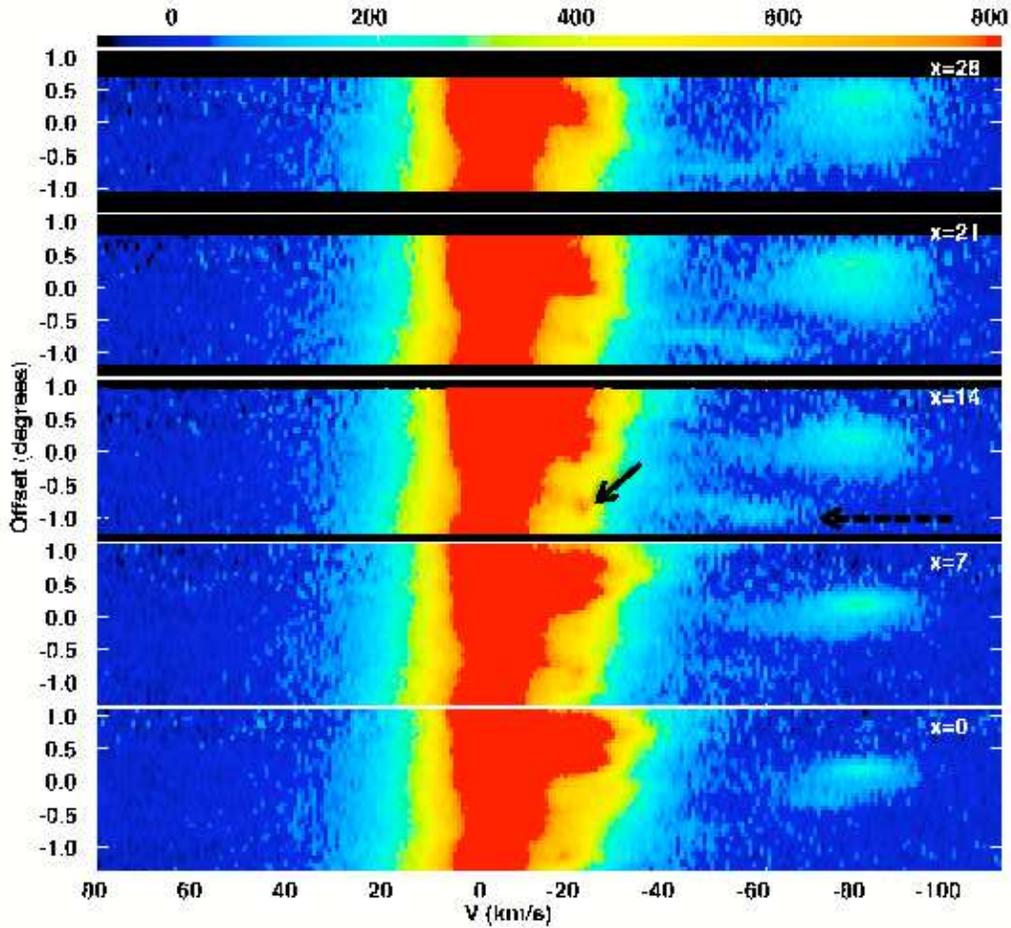}}}
\vspace{-7.0cm}
\caption{Position-velocity plots, derived at several positions
  parallel to the long axis of Cloud~I (PA=$309^{\circ}$;
  see Fig.~\ref{fig:OTFmom0}). 
The $y$ axis is the projected distance along the cloud
  in degrees, with values decreasing toward the southeast (i.e.,
  toward the ``tail'' of the cloud). The $x$-axis is the LSR velocity.
The solid arrow indicates the ``compact'' feature referred to in the
text (\S~\ref{GBTresults}), while the dashed arrow indicates an example of a
``streamer'' of material that
may have been ram pressure stripped from Cloud~I. 
The values in the upper right of 
each panel indicate the offset in arcminutes (along PA=309$^{\circ}$)
from the southwestern edge of the cloud. The intensity scale 
is in units of Jy pixel$^{-1}$.}
\label{fig:PVplots}
\end{figure}

\begin{figure}
\centering
\scalebox{0.9}{\rotatebox{0}{\includegraphics{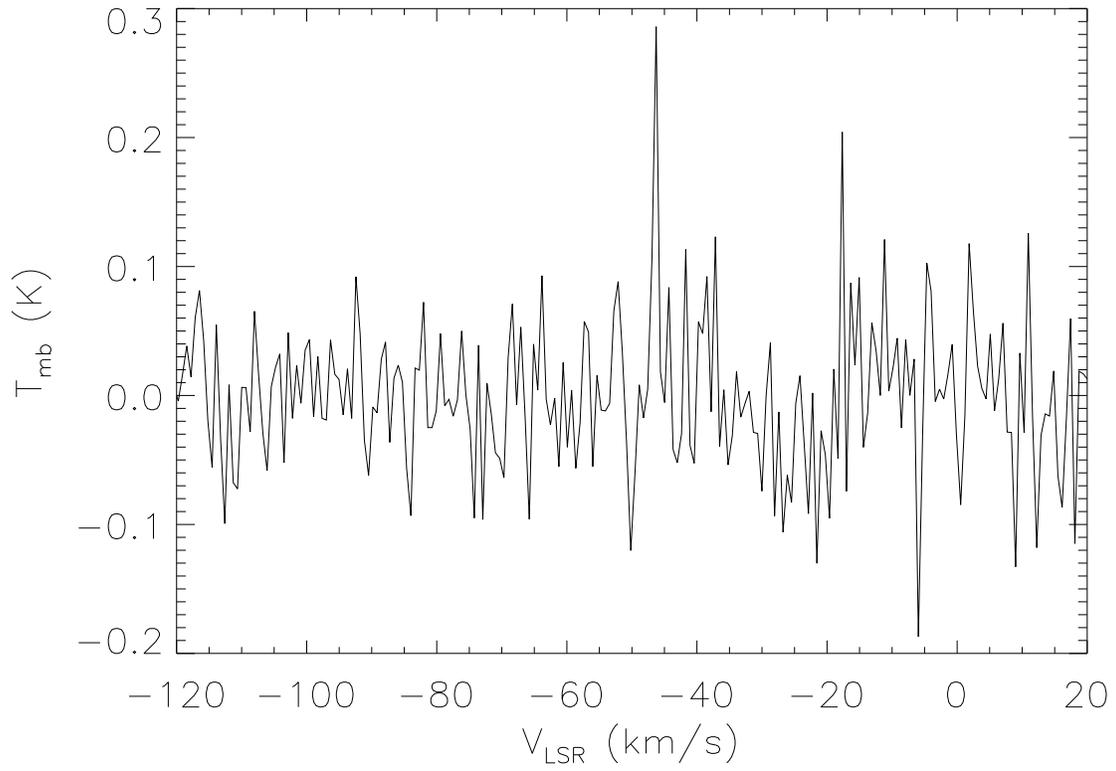}}}
\caption{$^{12}$CO $J$=1-0 spectrum toward the position of peak \HI\
  column density in Cloud~I. Nine frequency-switched
  spectra covering a 3$\times$3 beam grid with \am{8}{4} spacings,
  centered at $(l,b)=$(\ad{75}{92},\ad{+47}{78})  have
  been averaged (see \S~\ref{COobs}). }
\label{fig:COspectrum}
\end{figure}

\end{document}